\documentclass[11pt]{elsarticle}
\usepackage[top=2.0cm,bottom=2.5cm,left=2cm,right=2cm]{geometry}
\usepackage[utf8x]{inputenc}
\usepackage{helvet}
\usepackage{booktabs,nicefrac,microtype,natbib}
\usepackage{lineno,amsmath,color,soul,amsfonts}
\usepackage{float,graphicx,multirow,blindtext}
\usepackage[flushleft]{threeparttable}
\usepackage[outdir=./]{epsfig}
\modulolinenumbers[1]  
\usepackage{xfrac,bigints,enumitem,rotating}
\usepackage{relsize,longtable,mathtools,mathrsfs}
\usepackage[font=normalsize,labelfont={bf}]{caption}
\usepackage{siunitx}     
\usepackage{cancel}

\usepackage[colorinlistoftodos]{todonotes}
\usepackage{rotating}
\usepackage{pdflscape}
\usepackage{afterpage}
\usepackage{algpseudocode}
\usepackage{tikz}

\usepackage{esvect}

\usepackage{mathtools}
\DeclarePairedDelimiter\norm{\lVert}{\rVert}%

\usepackage[linesnumbered,ruled,longend]{algorithm2e}

\usepackage{esint}

\usepackage{tabularx, booktabs, threeparttable}

\makeatletter
\def\ps@pprintTitle{%
  \let\@oddhead\@empty
  \let\@evenhead\@empty
  \def\@oddfoot{\reset@font\hfil\thepage\hfil}
  \let\@evenfoot\@oddfoot
}
\makeatother

\usepackage{hyperref}
\hypersetup
    {
    colorlinks=true,
    linkcolor=blue,
    filecolor=magenta,     
    urlcolor=cyan
    }
\hypersetup{pdfauthor={Name}}

\allowdisplaybreaks

\begin{document}

\begin{frontmatter}

\title{Modelling Heterogeneous Interfaces using Element-based Finite Volumes}

\author[mymainaddress]{Suhaib Ardah\corref{mycorrespondingauthor}}
\cortext[mycorrespondingauthor]{Corresponding author: s.ardah19@imperial.ac.uk}
\author[mysecondaryaddress]{Francisco J. Profito}
\author[mymainaddress]{Daniele Dini}
\address[mymainaddress]{Department of Mechanical Engineering, Imperial College London, London, SW7 2AZ, UK}
\address[mysecondaryaddress]{Department of Mechanical Engineering, Polytechnic School of the University of São Paulo, São Paulo, Brazil}

\begin{abstract}
Accurately depicting multiphysics interactions in interfacial systems requires computational frameworks capable of reconciling geometric adaptability with strict conservation fidelity. However, traditional spatiotemporal discretisation methods often compromise between mesh flexibility and flow conservation enforcement, hence constraining their effectiveness in elucidating the underlying mechanisms. Here, we respond to these computational demands by developing a novel three-dimensional adaptation of the Element-based Finite Volume Method (EbFVM)---a hybrid numerical strategy that merges the geometric flexibility of Finite Element Methods with the conservation-centric principles of Finite Volume Methods. The proposed framework introduces advanced discretisation techniques tailored to unstructured, irregular mesh entities, including detailed parametric shape functions, robust flux integration schemes and rigorous body-fitted curvilinear coordinate mappings. Through a series of lubrication-driven benchmark problems, we demonstrate the EbFVM’s capacity to capture intricate transport phenomena, strong field couplings and scale disparities across geometrically complex domains. By enabling accurate modelling in geometrically and physically challenging interfacial systems, the three-dimensional EbFVM offers a versatile and generalisable tool for simulating transport phenomena in a plethora of multiphysics applications.
\end{abstract}


\end{frontmatter}

\section{Introduction}
Numerical discretisation provides a pathway for solving complex mathematical models which are otherwise analytically intractable in their continuous form. These models, typically described by partial differential equations (PDEs) \cite{Brezis1998Apr, Gershenfeld1998Nov, Farlow1994May}, underpin a wide array of physical and engineering phenomena by delineating the spatiotemporal evolution of multiphysical fields in applications ranging across fluid dynamics \cite{Fefferman2018Sep}, solid and contact mechanics \cite{Johnson1985}, heat and mass transfer \cite{Hatsopoulos1981}, electromagnetism \cite{Panofsky2012} and beyond. Nonetheless, deriving the analytical solution to these equations is often impractical in the presence of irregular geometries, pronounced non-linearities or complex boundary and initial conditions. Numerical solutions address these challenges by converting continuous physical domains into discrete counterparts such as grid points, finite elements or control volumes, thereby enabling the approximation of governing equations through discretisation methods \cite{Iserles2009}. This transformation enables the accurate numerical approximation of complex governing phenomena in engineering systems, where the levels of accuracy and computational efficiency are shaped by the choice of discretisation strategy, mesh resolution and the convergence performance of iterative solvers. Among the most prevalent discretisation techniques in computational mechanics are the Finite Difference, Finite Element and Finite Volume Methods, each offering distinct strengths suited to specific classes of problems. The application of a robust discretisation scheme is pivotal to ensuring the reliability and accuracy of numerical simulations.

Despite their widespread use, the aforementioned classical discretisation schemes exhibit inherent limitations when employed individually, particularly in scenarios involving complex geometries, steep field gradients or strict enforcement of conservation laws. The Finite Difference Method (FDM) provides numerical efficiency and algorithmic simplicity by leveraging Taylor series expansions \cite{Strikwerda2004Nov}; however, its dependence on the application of structured grids significantly limits its accuracy in irregular domains. In contrast, the Finite Element Method (FEM) offers superior geometric adaptability through its application of isoparametric mapping in unstructured meshes \cite{Zienkiewicz2013Sep}, which enables the accurate representation of complex boundaries. Nonetheless, it lacks inherent local conservation, making it less suitable for transport-dominated systems where flux balance is essential. The Finite Volume Method (FVM) overcomes this drawback by enforcing local conservation through flux-integrated formulations \cite{Moukalled}. However, traditional FVMs often struggle with geometric flexibility, especially near curved, evolving, or topologically complex boundaries and interfaces, leading to discretisation-induced errors in the solution of fine-scale features. These trade-offs underscore the need for hybrid numerical simulation frameworks that integrate the strengths of FDM, FEM, and FVM — namely, computational efficiency, flexibility, and conservation, respectively — to address the challenges posed by modern multiphysics systems.

The Element-Based Finite Volume Method (EbFVM) has emerged as a state-of-the-art hybrid-type numerical approach that coherently combines the geometric flexibility of FEM with the conservation-driven principles of FVM \cite{Honorio2018Jul, Xu2020}. Originally introduced under the name Control Volume Finite Element Method (CVFEM) \cite{Baliga1979, Baliga1980, Schneider1987}, the EbFVM overcomes the fundamental limitations of traditional methods, offering a unified framework that balances numerical accuracy, flexibility and flux conservation. In this formulation, the computational domain is discretised into finite elements, typically triangles, quadrilaterals, or their three-dimensional counterparts, enabling the solution to conform to complex geometries. Within these elements, interpolation and shape functions derived from FEM are utilised to approximate field variables, while local conservation laws are strictly enforced through embedded finite volume formulations constructed around nodal control volumes. These dual principles enable the accurate computation of transport fluxes across control volume boundaries, while ensuring local conservation throughout the computational domain \cite{Maliska2023}. By bridging the gap between the flexibility of FEM and the conservativeness of FVM, the EbFVM provides a robust and adaptable tool for solving advection–diffusion problems, making it particularly well-suited for multiphysics applications involving irregular domains and strongly coupled interfacial processes.

Although the EbFVM successfully integrates geometric adaptability with strong conservation principles, its application to multiphysics interfacial systems driven by fluid-solid interactions remains relatively unexplored. These systems are governed by complex couplings between distinct physical processes, such as fluid flow, solid deformation, heat and mass transfer, each characterised by its governing equations, spatial scales and temporal dynamics \cite{Rugonyi2001, Tallec2001}. Multiphysics interactions of this nature are ubiquitous across a broad spectrum of applications, spanning engineering systems such as sliding and rolling bearings \cite{Hamrock2004Mar}, bridge structures \cite{Wuchner2007Jun}, aerospace assemblies \cite{Farhat2006Mar} and hydraulic actuators \cite{Peixin2021}, as well as biological systems including arterial blood flow \cite{Bazilevs2006Sep} and cartilage rehydration dynamics \cite{Putignano2021Apr}. Capturing the full complexity of these phenomena presents formidable numerical challenges due to the strong coupling between time–dependent advection–diffusion–reaction processes, transient interfacial stresses and the continuously evolving material boundaries. These systems often exhibit abrupt, multi-scale feedback, such as severe pressure gradients in ultra-thin lubrication films, permeability heterogeneities in porous structures, or rapid temperature variations in phase-changing flows, all of which can limit the performance of traditional discretisation techniques. Numerous standard numerical approaches lack either the ability to resolve complex structures or impose local and global flow conservation in the discrete formulation. Hence, devising a computational framework that can simultaneously tackle sharp gradients, maintain conservation laws, accommodate complex geometries and achieve numerical stability is paramount.

These challenges are addressed in the present contribution by developing a novel three-dimensional generalisation of the Element-Based Finite Volume Method (EbFVM) framework, which utilises a body-fitted curvilinear coordinate transformation to solve multiphysics problems at heterogeneous interfaces. Through the deployment of body-fitted curvilinear coordinate transformations, the EbFVM enables precise mapping from non-orthogonal, curvilinear grids in the physical domain to structured, orthogonal grids in the computational domain, thereby enhancing geometric adaptability and numerical accuracy. The proposed methodology employs both hexahedral and triangular prism elements for spatial discretisation to accommodate the geometrical complexities inherent in real-world engineering and scientific applications. To rigorously validate the versatility, robustness and accuracy of the proposed interfacial EbFVM, lubrication problems are employed as benchmark cases, driven by the multiphysics challenges involving thin-film hydrodynamics, interfacial contact interactions and transport processes where flow conservation is vital \cite{Vakis2018Sep, Ardah2025}. Moreover, they exhibit extreme length scale disparities between film thickness and domain extent, strong coupling between fluid flow and surface deformation, as well as sharply varying pressure and temperature fields. By addressing these benchmark scenarios, the proposed EbFVM framework not only highlights its superior conservation properties and geometric adaptability but also reveals the shortcomings of traditional methods when applied to complex, multiphysics interfacial systems.

The remainder of this work is organised as follows. \autoref{sec:EbFVM_Overview} introduces the Element-Based Finite Volume Method, outlining its fundamental principles, historical emergence and development into a robust hybrid numerical framework. \autoref{sec:EbFVM_Disretisation} explicates the spatial and temporal discretisation strategies used to solve governing transport equations, with particular emphasis on element-wise formulations, flux integration techniques and the assembly of the global system of linear equations. \autoref{sec:Benchmark} presents the governing equations and numerical simulation setup employed to evaluate the proposed approach, providing the foundation for a sequence of lubrication-based benchmark problems discussed in \autoref{sec:Results}, which serve to assess the method’s accuracy, robustness and effectiveness in capturing interfacial transport phenomena. \autoref{sec:Conclusion} explores the extensibility of the EbFVM, highlighting its applicability to a wide range of transport-driven systems characterised by geometric complexity and multiphysics coupling, while also encapsulating the broader contributions and significance of the proposed methodology.

\section{Fundamentals of the Element-based Finite Volume Method}
\label{sec:EbFVM_Overview}
The Element-Based Finite Volume Method (EbFVM) has gained recognition as a robust and versatile numerical strategy for simulating conservation-governed systems across a broad spectrum of engineering applications, including fluid dynamics \cite{Prakash1985, Honorio2018Jul}, heat and mass transfer \cite{Baliga1979, Patankar1980Jan} and reservoir modelling \cite{Forsyth1990, Hurtado2007Apr}. Evolving from the early concept of the Control Volume Finite Element Method (CVFEM) \cite{Baliga1980, Raw1985, Rousse1991}, the EbFVM bridges the methodological gap between the geometric flexibility provided by finite element discretisation and the strict local conservation offered by finite volume schemes. Its early success in reservoir simulations and porous media applications \cite{Gottardi1992, Verma1997, Prevost2001, Marcondes2007} showcases its ability to handle spatial heterogeneities and irregular geometries with numerical efficiency. However, early formulations exhibited inconsistencies when applied to multiphase scenarios, where oversimplified assumptions and coarse approximations led to the introduction of non-physical artefacts \cite{Cordazzo2004a, Cordazzo2004b}. These shortcomings catalysed further methodological advancements aimed at enhancing the robustness and accuracy of the approach for multiphysics transport problems.

\begin{figure}[H]
   \centering
   \includegraphics[width=\linewidth]{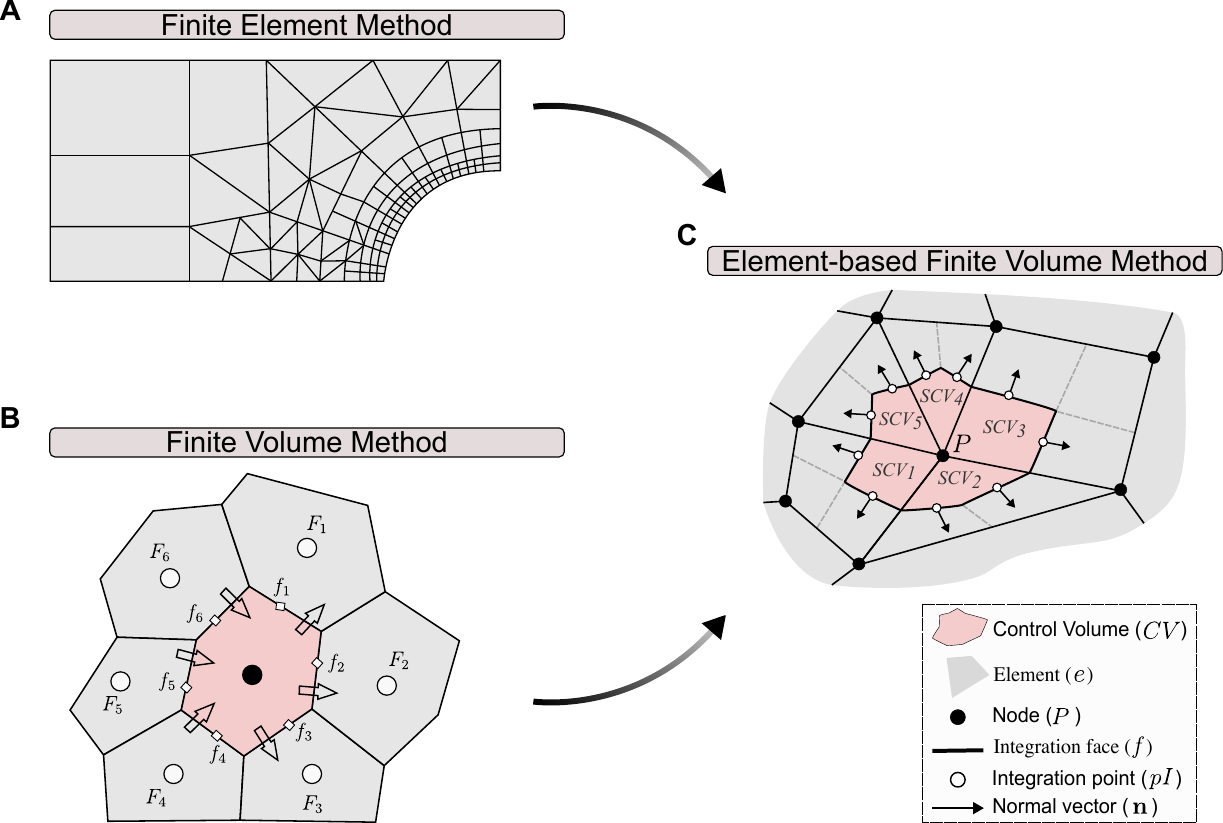}
   \caption{Overview of the Element-Based Finite Volume Method (EbFVM). The EbFVM synergistically integrates: (\textbf{A}) the geometric adaptability of the Finite Element Method (FEM), enabling accurate conformity to complex domains, and (\textbf{B}) the conservation principles of the Finite Volume Method (FVM), ensuring rigorous balance of transport fluxes across control volumes. (\textbf{C}) A schematic representation of the key geometric entities that form the foundation of the EbFVM framework.}
   \label{fig:EbFVM_Overview}
\end{figure}

The numerical implementation of the EbFVM begins with the discretisation of the physical domain into a mesh of simple-shaped elements ($e$), interconnected via node points ($P$) positioned at the element vertices. This ensemble of elements and nodes defines the \emph{geometric grid}, which forms the structural foundation for all subsequent numerical operations, as illustrated in \autoref{fig:EbFVM_Overview}. In practical applications, the mesh may consist of triangular or quadrilateral elements in two-dimensional space, and tetrahedral, hexahedral, or prismatic elements in three-dimensional domains. Notably, these elements do not serve directly as control volumes (CVs); instead, they provide the geometric scaffold from which the computational CVs are systematically constructed. This deliberate decoupling between the geometric and computational grids imparts a high degree of flexibility to the EbFVM, enabling accurate representation of complex geometries and spatial heterogeneities without compromising the rigorous enforcement of local and global conservation laws. As a result, the method is particularly well-suited for modelling transport phenomena in systems characterised by anisotropy, heterogeneity and intricate boundary features.

Departing from traditional FVMs that adopt a cell-centred approach, the EbFVM employs a cell-vertex formulation whereby elements are treated as auxiliary geometric entities rather than primary computational cells. In this approach, CVs are constructed around nodes rather than elements, yielding a secondary computational grid that facilitates the enforcement of local conservation laws \cite{Profito2015Sep}. As shown in \autoref{fig:EbFVM_Overview}\textbf{C}, in two-dimensional domains, each CV is associated with a node ($P$) and is composed of portions of the neighbouring elements known as sub-control volumes (SCVs). These SCVs are defined through a barycentric subdivision (or median rule), where lines are drawn from the element centroid to the midpoints of its edges, thereby forming CV faces, denoted as $f$, which serve as the fundamental entities for flux evaluation. The midpoint rule is then applied to approximate fluxes across these faces, ensuring both numerical accuracy and stability. Integration points ($pI$) are located at the centroids of each face, with corresponding normal vectors ($\boldsymbol{\mathrm{n}}$) defining the face direction. Through this subdivision scheme, each CV associated with a node ($P$) can be interpreted as an aggregate of its neighbouring SCVs, with fluxes at each integration point computed based on local geometric and material properties of the element in which it resides \cite{Hurtado2011, Hurtado2012}. Such formulation allows the EbFVM to effectively resolve conservation-driven transport phenomena over unstructured grids, ensuring accurate representation of intricate geometrical features.

In the cell-vertex formulation adopted by the EbFVM, the primary unknowns, such as pressure, velocity and temperature fields, are evaluated at the nodes of the geometric grid rather than at element centres, as is common in traditional FVMs. To account for the geometric distortions inherent in irregular meshes and to capture localised variations in transport properties, the EbFVM utilises predefined families of interpolation and shape functions. These functions are defined in a local (transformed) coordinate system, allowing calculations to be performed in a reference domain where integration becomes more tractable and numerically stable \cite{Profito2015Sep}. This transformation not only simplifies the computational implementation but also enhances the numerical accuracy when resolving sharp gradients in complex geometries. Once the integration points within each element are identified, the fluxes of transport properties across the SCV faces are evaluated using the nodal geometrical and material attributes. By considering each element as the fundamental computational unit, fluxes can be independently calculated and then assembled into a global system of equations using an element-by-element approach, similar to the strategies employed in the FEM. However, unlike the FEM, the EbFVM preserves the flow conservation that the FVM guarantees. Such a hybrid assembly process ensures that local interactions are consistently reconciled with global conservation, yielding stable, physically accurate solutions across arbitrarily shaped domains \cite{Maliska2023}.

The extension of the EbFVM from two-dimensional (2D) to three-dimensional (3D) domains introduces substantial computational challenges. These arise not only from the increased geometric complexity and higher degrees of freedom, but also from the need for more sophisticated interpolation and integration schemes to preserve numerical accuracy across volumetric elements \cite{Hurtado2011}. In the present work, which is particularly focused on interfacial problems, the transition to 3D is structured along the vertical direction, following a layered topology that enables systematic construction of prismatic elements. This approach mitigates the complete irregularity typically associated with fully unstructured 3D meshes, thereby enhancing computational tractability without sacrificing geometric fidelity. The move to three-dimensional domains substantially expands the applicability of the EbFVM to more realistic interfacial multiphysics systems, where spatial gradients and coupled phenomena evolve in three-dimensional spaces.

\section{Numerical Discretisation of the Element-based Finite Volume}
\label{sec:EbFVM_Disretisation}
This section outlines the three-dimensional formulation and discretisation strategy of the EbFVM, using the general conservation law for advection–diffusion processes as a representative example. Expressed in its strong conservative form, \autoref{eq:Advection-Diffusion} serves as the foundational equation for illustrating the spatiotemporal discretisation approach adopted in the proposed framework, given as:
\begin{align}
        \frac{\partial}{\partial t} \bigg( \rho {\phi} \bigg) + \boldsymbol{\nabla} \cdot \bigg( \rho \boldsymbol{v} {\phi} \bigg) = \boldsymbol{\nabla} \cdot \bigg( \boldsymbol{\Gamma}^{\phi} \boldsymbol{\nabla} {\phi} \bigg) + {Q^{\phi}}.
        \label{eq:Advection-Diffusion}
\end{align}
In the above equation, ${\phi}$ denotes the transport scalar quantity (\emph{e.g.} temperature or concentration), $\rho$ is the fluid density, $\boldsymbol{v}$ is the velocity field, $\boldsymbol{\Gamma}^{\phi}$ is the diffusion coefficient tensor and ${Q^{\phi}}$ is the source or sink term. This equation underpins a wide range of engineering problems governed by interfacial fluid–solid interactions, from heat and mass transfer in complex domains to turbulence modelling, reactive flows and other multiphysics phenomena.

The methodology presented in this study builds upon the foundational work in \cite{Profito2015Sep, Profito2015, Profito2019Dec}, where the EbFVM approach was initially introduced in a two-dimensional framework to investigate the lubrication behaviour of connecting rod bearings of internal combustion engines, and the influence of surface texturing on the lubrication performance of rubbing contacts under isothermal conditions. These early developments demonstrated the method’s capability to efficiently resolve lubrication problems with fluid film cavitation in complex geometries while enforcing flow conservation. Building on this groundwork, the present contribution extends the EbFVM into three dimensions, thereby enhancing its scope to encompass multiphysics interfacial problems involving fluid–solid interactions, heat transfer, and evolving domain topologies. The proposed three-dimensional formulation retains the method’s core strengths — geometric adaptability and strict conservation enforcement — while extending its relevance to a broader class of transport-driven systems.

\subsection{Three-Dimensional Geometric Entities}
\begin{figure}[H]
   \centering
   \includegraphics[width=0.95\linewidth]{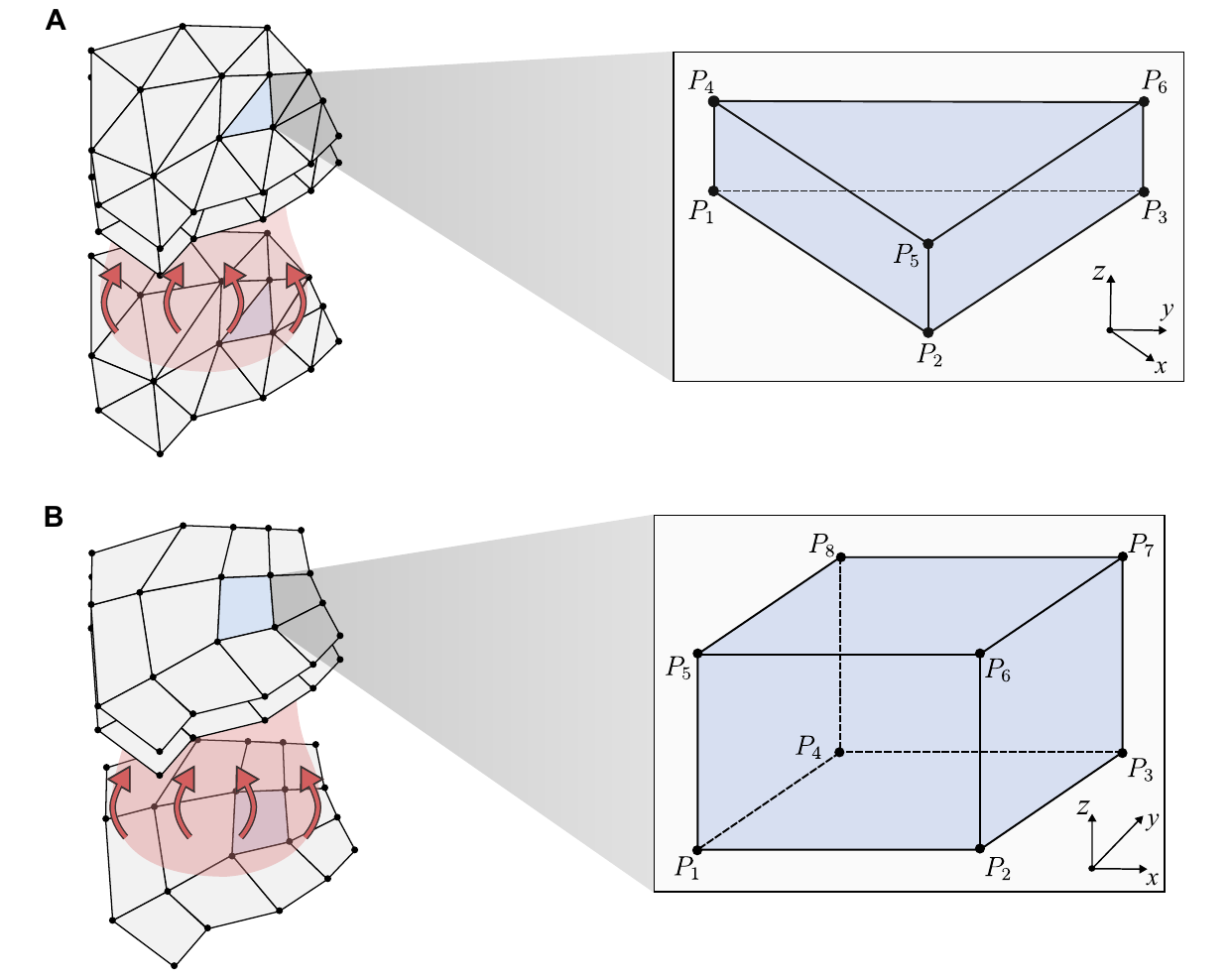}
   \caption{Three-dimensional geometric elements employed in the proposed EbFVM discretisation framework. (\textbf{A}) Triangular prism elements with six nodes, offering enhanced flexibility for conforming to irregular or curved geometries. (\textbf{B}) Hexahedral elements with eight nodes, facilitating structured meshing and efficient flux integration.}
   \label{fig:TriangularPrism_Hexahedral}
\end{figure}

The computational domain is discretised using three-dimensional geometric elements, specifically six-node triangular prisms and eight-node hexahedral elements, as depicted schematically in \autoref{fig:TriangularPrism_Hexahedral}\textbf{A} and \autoref{fig:TriangularPrism_Hexahedral}\textbf{B}, respectively. Triangular prism elements, featuring triangular bases connected through three quadrilateral faces, offer enhanced adaptability to intricate and curved boundaries, making them particularly effective in domains defined by highly irregular geometries. On the other hand, hexahedral elements, comprising six quadrilateral faces, twelve edges and eight vertices, support a more structured grid arrangement for accurate flux evaluation and efficient numerical integration. As previously discussed, by incorporating both element types, the EbFVM framework enables more precise representation of complex geometries while maintaining rigorous flux conservation and accurate resolution of spatial gradients in multiphysics environments. Comprehensive information on the shape function formulation and coordinate transformation strategies for these element types can be found in \ref{sec:EbFVM_Discretisation}.

\subsection{Temporal Discretisation}
The temporal discretisation of the general transport equation (\autoref{eq:Advection-Diffusion}) is performed by integrating all terms over a finite time interval \(\Delta t\), spanning from the current time level \(t\) to the subsequent time level \(t + \Delta t\). This procedure yields the following temporally integrated equation:
\begin{equation}
        \bigintsss_{t}^{t+\Delta t} \left[ \frac{\partial}{\partial t} \Big( \rho {\phi} \Big) \right] dt 
        +
        \bigintsss_{t}^{t+\Delta t} \bigg[ \boldsymbol{\nabla} \cdot \Big( \rho \boldsymbol{v} {\phi} \Big) \bigg] dt 
        =
        \bigintsss_{t}^{t+\Delta t} \bigg[ \boldsymbol{\nabla} \cdot \Big( \boldsymbol{{\Gamma}}^{\phi} \boldsymbol{\nabla} {\phi} \Big) \bigg] dt 
        +
        \bigintsss_{t}^{t+\Delta t} \bigg[ {Q^{\phi}} \bigg] dt.
        \label{eq:1}
\end{equation}

To maintain the numerical stability and robustness of the solution, the first-order implicit backward Euler method is adopted for time integration. The primary benefit of this formulation lies in its unconditional stability, ensuring that the numerical accuracy of the solution is predominantly influenced by temporal resolution governed by the time step \(\Delta t\). Discretising \autoref{eq:1} at discrete time levels $t = n \Delta t$, where $ n$ denotes the $n$-th time step, results in the following semi-discrete formulation:
\begin{equation}
        \bigg[ \frac{({\rho\phi})^{n} - ({\rho\phi})^{n-1}}{\Delta t} \bigg] + \bigg[ \boldsymbol{\nabla} \cdot \Big( \rho \boldsymbol{v} {\phi} \Big) \bigg]^{n} = \bigg[ \boldsymbol{\nabla} \cdot \Big( \boldsymbol{{\Gamma}}^{\phi}\boldsymbol{\nabla} {\phi} \Big) \bigg]^{n} + \bigg( {Q^{\phi}} \bigg)^{n}.
        \label{eq:Transient_Equation}
\end{equation}

\subsection{Spatial Discretisation}
To discretise the spatial domain in \autoref{eq:Transient_Equation}, the governing advection–diffusion equation is first reformulated into its integral (weak) form. This transformation is commonly achieved through the Weighted Residual Method (WRM) \cite{Hutton2004}, which systematically reduces approximation errors by requiring the residuals of the discretised equations to be orthogonal to a set of weighting functions. By doing so, the WRM reduces approximation errors and promotes both numerical stability and physical consistency, particularly in the enforcement of conservation laws over complex, irregular geometries \cite{Reddy2005}.

Applying the WRM, the strong form of the transport equation is integrated over the entire computational domain \({\widehat{\Omega}}\), yielding the following expression:
\begin{equation}
        \bigintsss \!\!\!\! \bigintsss \!\!\!\! \bigintsss_{{\widehat{\Omega}}} 
        \mathcal{W} \Bigg\{
            \bigg[ \frac{({\rho\phi})^{n} - ({\rho\phi})^{n-1}}{\Delta t} \bigg] + \bigg[ \boldsymbol{\nabla} \cdot \Big( \rho \boldsymbol{v} {\phi} \Big)^{n} \bigg] - \bigg[ \boldsymbol{\nabla} \cdot \Big( \boldsymbol{{\Gamma}}^{\phi} \boldsymbol{\nabla} {\phi} \Big)^{n} \bigg] - \bigg( {Q^{\phi}} \bigg)^{n}
        \Bigg\} \; d{\widehat{\Omega}} = 0,
        \label{eq:SpatialDisc_V1}
\end{equation}
where $\mathcal{W}$ is a weighting function defined over ${\widehat{\Omega}}$. In the context of finite volume methods, the subdomain weighting approach is adopted \cite{Lohner2008}, where $\mathcal{W}$ is taken to be unity within each control volume and zero elsewhere. This localises the weak formulation to node-centred subdomains, allowing the governing equations to be evaluated independently over each control volume, leading to the following discretised integral expression:
\begin{equation}
    \begin{split}
        \bigintsss \!\!\!\! \bigintsss \!\!\!\! \bigintsss_{{\Omega}} 
        \bigg[ \frac{({\rho\phi})^{n} - ({\rho\phi})^{n-1}}{\Delta t} \bigg] d{\Omega} 
        &+ \bigintsss \!\!\!\! \bigintsss \!\!\!\! \bigintsss_{{\Omega}} 
        \bigg[ \boldsymbol{\nabla} \cdot \Big( \rho \boldsymbol{v} {\phi} \Big)^{n}  \bigg] d{\Omega} \\
        &= \bigintsss \!\!\!\! \bigintsss \!\!\!\! \bigintsss_{{\Omega}} 
        \bigg[ \boldsymbol{\nabla} \cdot \Big( \boldsymbol{{\Gamma}}^{\phi} \boldsymbol{\nabla} {\phi} \Big)^{n} \bigg] d{\Omega} + \bigintsss \!\!\!\! \bigintsss \!\!\!\! \bigintsss_{{\Omega}} 
        \bigg( {Q^{\phi}} \bigg)^{n} d{\Omega},
    \end{split}
    \label{eq:Eq_a}
\end{equation}
where ${\Omega}$ denotes the control volume associated with node $P$.

To convert the volume integrals of the convective and diffusive terms in \autoref{eq:Eq_a} into more convenient surface integrals, the divergence (or Gauss–Ostrogradsky's) theorem is applied. This yields the following integral form over the control volume boundary ${\vartheta}$:
\begin{equation}
    \begin{split}
        \bigintsss \!\!\!\! \bigintsss \!\!\!\! \bigintsss_{{\Omega}} 
        \bigg[ \frac{({\rho\phi})^{n} - ({\rho\phi})^{n-1}}{\Delta t} \bigg] d{\Omega} &+ 
        \mathlarger{\mathlarger{\oiint}}_{{\vartheta}} 
        \bigg[ \Big( \rho \boldsymbol{v} {\phi} \Big)^{n} \cdot {\boldsymbol{\mathrm{n}}} \bigg] d{\vartheta} \\
        &= \mathlarger{\mathlarger{\oiint}}_{{\vartheta}} 
        \bigg[ \Big( \boldsymbol{{\Gamma}}^{\phi} \boldsymbol{\nabla} {\phi} \Big)^{n} \cdot {\boldsymbol{\mathrm{n}}} \bigg] d{\vartheta} + \bigintsss \!\!\!\! \bigintsss \!\!\!\! \bigintsss_{{\Omega}} 
        \bigg( {Q^{\phi}} \bigg)^{n} d{\Omega},
    \end{split}
    \label{eq:Eq1}
\end{equation}
where ${\boldsymbol{\mathrm{n}}}$ denotes the outward-pointing unit normal vector on each control volume face, as illustrated in \autoref{fig:EbFVM_Overview}\textbf{C}.

Subsequently, the fluxes across control volume faces are approximated using the second-order midpoint rule, in which each surface integral is evaluated as the product of the integrand at the face midpoint and the corresponding face area, whereas each volume integral is computed as the product of the integrand at the CV centroid and the corresponding CV volume \cite{Ferziger}. Applying this to a node-centred control volume leads to the following discrete formulation:
\begin{equation}
    \begin{split}
        \Bigg[
           \frac{(\rho{\phi})^n}{\Delta t}
        \Bigg]_{P}
        \underbrace{\mathlarger{\mathlarger{\sum}}_{e \in \mathcal{E}_P} \Big[ \Delta \Omega^{e}_{s} \Big]}_{\Delta \Omega_{P}}
        &-
        \Bigg[
           \frac{(\rho{\phi})^{n-1}}{\Delta t}
        \Bigg]_{P} \underbrace{\mathlarger{\mathlarger{\sum}}_{e \in \mathcal{E}_P} \Big[ \Delta \Omega^{e}_{s} \Big]}_{\Delta \Omega_{P}}
        +
        \mathlarger{\mathlarger{\sum}}_{e \in \mathcal{E}_P} 
        \mathlarger{\mathlarger{\sum}}_{f \in \mathcal{F}_P^e} 
        \bigg[
            \Big( \rho \boldsymbol{v} {\phi} \Big)^{n} \cdot {\boldsymbol{\mathrm{n}}}^*
        \bigg]^{e}_{f} \\
        &=
        \mathlarger{\mathlarger{\sum}}_{e \in \mathcal{E}_P} 
        \mathlarger{\mathlarger{\sum}}_{f \in \mathcal{F}_P^e} 
        \bigg[
            \Big( \boldsymbol{{\Gamma}}^{\phi} \boldsymbol{\nabla} {\phi} \Big)^{n} \cdot {\boldsymbol{\mathrm{n}}}^*
        \bigg]^{e}_{f} 
        +  
        \bigg[ 
            {Q^{\phi}}
        \bigg]^{n}_P
        \underbrace{\mathlarger{\mathlarger{\sum}}_{e \in \mathcal{E}_P} \Big[ \Delta \Omega^{e}_{s} \Big]}_{\Delta \Omega_{P}},
        \label{eq:Eq_10}
    \end{split}
\end{equation}
where $\mathcal{E}_P$ denotes the set of elements surrounding node $P$ and that collectively define its CV, and $\mathcal{F}_P^e$ the faces within each element $e$ contributing to the CV flux balance. The term ${\boldsymbol{\mathrm{n}}_f^*}^e = {\boldsymbol{\mathrm{n}}}_f^e \Delta {\vartheta^e_f}$ embodies both the orientation (unit normal vector ${\boldsymbol{\mathrm{n}}}_f^e$) and area $\Delta \vartheta_f^e$ of the CV face. The quantity \(\Delta \Omega^{e}_{s}\) represents the volume of the SCV within element \(e\) contributing to the nodal CV, and the total control volume around node $P$ is denoted by \(\Delta \Omega_P\). A detailed account of SCV construction and element partitioning is provided in \ref{sec:EbFVM_Discretisation}, where the geometric and topological foundations of the EbFVM are formally developed.

The fully discretised conservation law for each node-centred control volume is encapsulated in \autoref{eq:Eq_10}. Within this formulation, the inner summations over element faces represent local contributions from convective and diffusive transport across individual SCVs, while the outer summations ensure global conservation over the entire nodal control volume. The global system of discrete equations is assembled element-wise using a parametric integration framework that accommodates arbitrary element orientation, distortion, and positioning. This approach supports the systematic generation of element-level matrices and vectors, echoing the modularity and structure of the finite element assembly process.

\subsection{Approximation of Diffusive Fluxes}
The contribution of diffusive fluxes across all SCV faces associated with a given element \(e\), which corresponds to the fourth term in \autoref{eq:Eq_10}, can be expressed in compact matrix form as:
\begin{equation}
        \boldsymbol{\mathfrak{D}}_e^{n} = \boldsymbol{\mathcal{D}}_e^{n} \, {\boldsymbol{\phi}_e^{n}},
\end{equation}
where \(\boldsymbol{\mathfrak{D}}_e^{n}\) denotes the vector of discrete diffusive flux contributions through each SCV within element \(e\) and \(\boldsymbol{\phi}_e^{n}\) contains the nodal values of the transport scalar field \(\phi\). The element diffusivity matrix \(\boldsymbol{\mathcal{D}}_e^{n}\) encapsulates the influence of the element’s geometry, material diffusivity and spatial discretisation. This matrix serves as a key building block in enforcing local diffusion balances over arbitrarily shaped elements. A detailed derivation of \(\boldsymbol{\mathcal{D}}_e^{n}\) is provided in \ref{sec:Appendix_Diffusion}.

\subsection{Approximation of Convective Fluxes}
Analogous to the treatment of diffusive transport, the third term in \autoref{eq:Eq_10} captures the net convective flux across all SCVs associated with a given element \(e\) in the three-dimensional computational domain. This term can be expressed in compact matrix form as follows:
\begin{equation}
        \boldsymbol{\mathfrak{C}}_e^{n} = \boldsymbol{\mathcal{C}}_e^{n} \, \boldsymbol{\phi}_e^{n},
\end{equation}
where \(\boldsymbol{\mathfrak{C}}_e^{n}\) denotes the vector of discrete convective flux contributions through each SCV within element \(e\) and \(\boldsymbol{\phi}_e^{n}\) contains the nodal values of the transport scalar field $\phi$. The matrix \(\boldsymbol{\mathcal{C}}_e^{n}\), referred to as the element convection matrix, encapsulates the influence of the local velocity field and the element's geometric configuration on the convective fluxes. It acts as the directional transport operator within the control volume context, accounting for the orientation, flow alignment and local topological distortion of the element. A detailed derivation of \(\boldsymbol{\mathcal{C}}_e^{n}\), including its construction based on interpolation schemes, velocity projections and integration strategies, is provided in \ref{sec:Appendix_Convection}.

\subsection{Approximation of the Source Terms}
The contributions arising from the temporal and volumetric source terms, corresponding to the first, second and fifth terms in \autoref{eq:Eq_10}, are evaluated over the SCVs within each element \(e\). Their discrete representations are expressed as:
\begin{subequations}
    \begin{align}
        \boldsymbol{\mathfrak{S}}_{1,e}^{n} &= \boldsymbol{\mathcal{S}}_{1,e}^{n} \cdot \boldsymbol{\phi}_{e}^{n}, \\
        \boldsymbol{\mathfrak{S}}_{2,e}^{n} &= \boldsymbol{\mathcal{S}}_{2,e}^{n} \cdot \boldsymbol{\phi}_{e}^{n-1}, \\
        \boldsymbol{\mathfrak{S}}_{3,e}^{n} &= \boldsymbol{\mathcal{S}}_{3,e}^{n},
    \end{align}
\end{subequations}
where \(\boldsymbol{\mathfrak{S}}_{1,e}^{n}\), \(\boldsymbol{\mathfrak{S}}_{2,e}^{n}\) and \(\boldsymbol{\mathfrak{S}}_{3,e}^{n}\) denote the discrete nodal source vectors for element \(e\) corresponding to, respectively, the transient term at time level $n$, the historical contribution from time level $n-1$, and the volumetric source term. The auxiliary vectors \(\boldsymbol{\mathcal{S}}_{1,e}^{n}\), \(\boldsymbol{\mathcal{S}}_{2,e}^{n}\) and \(\boldsymbol{\mathcal{S}}_{3,e}^{n}\) incorporate the geometric, physical and material properties required for the integration over each SCV and are computed using quadrature schemes. The vectors \(\boldsymbol{\phi}_{e}^{n}\) and \(\boldsymbol{\phi}_{e}^{n-1}\) contain the nodal values of the transport scalar quantity at the current and previous time steps, respectively. The operator “\(\cdot\)” denotes the standard Euclidean inner product. Additional derivation details, including quadrature rule selection and implementation nuances, are provided in \ref{sec:Appendix_SourceTerm}.

\subsection{Assembly of the Global System of Equations}
Following the element-wise calculation of the diffusivity and convective matrices and source term vectors, the local contributions are assembled into a global system of discrete equations. This process combines all element-level quantities into CVs centred at computational nodes $P$, ensuring both local and global conservation principles throughout the solution domain. The resulting global system of equations ensures that the total flux balance, including advection, diffusion, and source effects, is maintained at each nodal control volume. The discrete transport equation at node $P$ can be expressed in a general form as follows:
\begin{equation}
        \Big(a_P^n\Big)\phi_P^n 
        + 
        \mathlarger{\mathlarger{\mathlarger{\sum}_{Q \in \mathcal{N}_P}}}
        \bigg[
            \Big(a_{PQ}^n + b_{PQ}^n\Big)\,\phi_Q^n
        \bigg]
        = 
        B_P^n,
        \label{eq:GlobalSystem}
\end{equation}
where $\phi_P^n$ and $\phi_Q^n$ denote the values of the transported variable at node $P$ and its neighbouring nodes $Q \in \mathcal{N}P$ at the current time level $n$, respectively. The diagonal coefficient $a_P^n$ represents the combined temporal contribution at node $P$, while $a_{PQ}^n$ and $b_{PQ}^n$ denote the diffusive and convective coupling coefficients between nodes $P$ and $Q$, respectively. The term $B_P^n$ aggregates source and non-linear contributions, including effects arising from external forces or prescribed boundary conditions. This formulation explicitly separates diffusive and convective interactions, thereby preserving both local and global conservation within the element-based finite volume framework. By systematically enforcing this conservative balance at each node, the fully discretised transport equation results in the following global system of linear equations:
\begin{equation}
        \boldsymbol{\mathrm{A}}^n \boldsymbol{\phi}^n = \boldsymbol{\mathrm{b}}^n,
\end{equation}
where $\boldsymbol{\mathrm{A}}^n$ is the global coefficient matrix, $\boldsymbol{\phi}^n$ is the vector of unknown nodal values and $\boldsymbol{\mathrm{b}}^n$ is the assembled right-hand side vector containing source contributions. This element-wise assembly strategy retains the geometric adaptability and conservation-oriented nature of the EbFVM, achieving computational stability and physical accuracy even in the presence of highly irregular geometries, distorted meshes and complex multiphysics couplings.

\section{Benchmark Application: Multiphysics Lubrication Interfaces}
\label{sec:Benchmark}
\begin{figure}[H]
        \centering
        \includegraphics[width=0.95\linewidth]{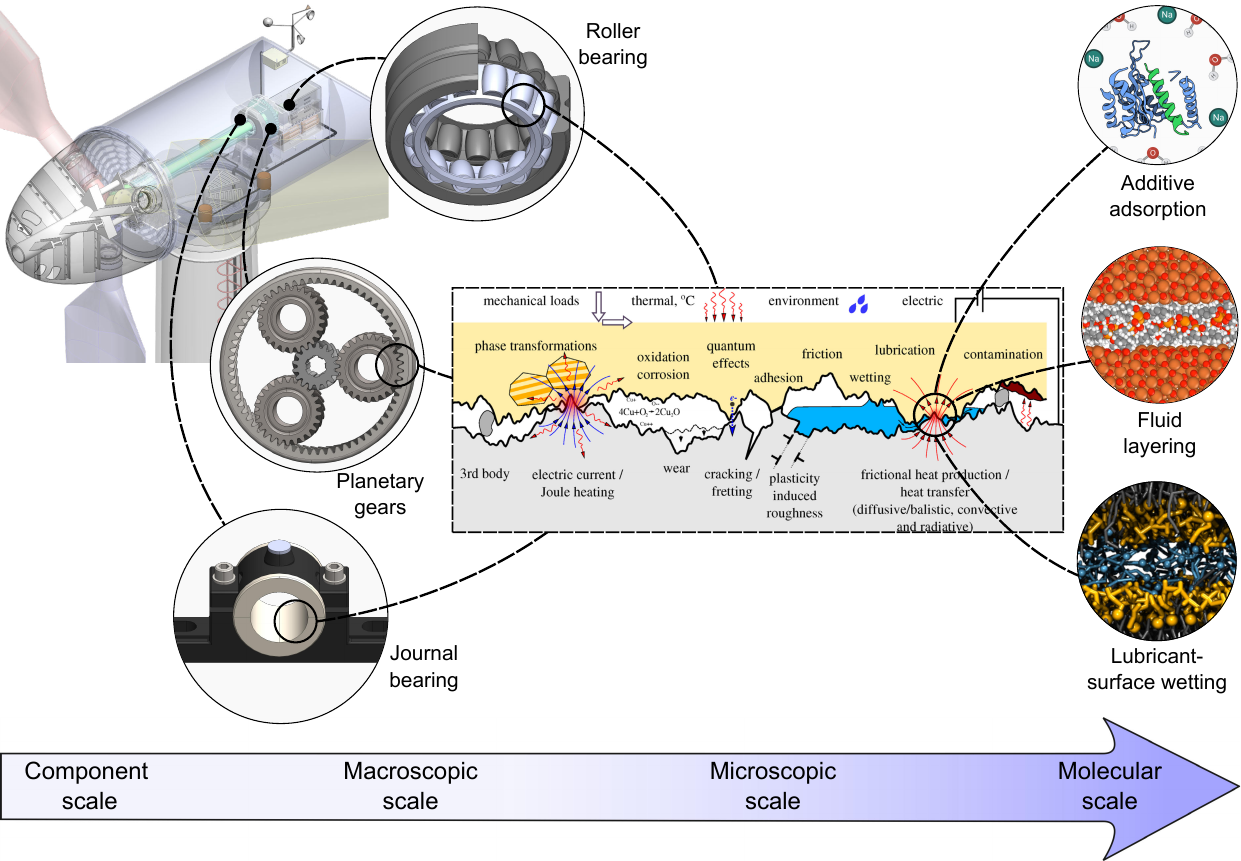}
        \caption{The multiphysics and multiscale nature of lubricated interfaces. Illustration of the hierarchical transition from component-level systems (\emph{e.g.}, wind turbine assemblies) to macro-, micro- and molecular-scale levels, capturing the escalating complexity of interfacial phenomena. As spatial scales descend, the interplay of the physicochemical interactions becomes increasingly intricate. Microscopic schematic is adapted from {\cite{Vakis2018Sep}}, with permission from Elsevier.}
        \label{fig:Tribological_Interactions}
\end{figure}

Lubricated interfaces embody a compelling and challenging class of multiphysics interfacial problems, offering a physically ideal benchmark to assess the capabilities of the proposed EbFVM approach. Within the broader field of tribology, the science of interacting surfaces in relative motion, lubrication systems encompass a wide spectrum of coupled fluid and solid thermomechanical phenomena. These include laminar viscous flow, fluid rheology and thermally induced property variations, and elastic or plastic deformation of bounding solids, all evolving concurrently across different spatial and temporal scales \cite{Khonsari2017Aug, Szeri2010Dec}, as depicted in \autoref{fig:Tribological_Interactions}. At the core of the lubrication theory \cite{Frene1997Nov} are the dynamic interactions between the thin lubricating fluid film and the surrounding contacting surfaces, representing a fluid–structure interaction (FSI) problem \cite{Bungartz2007}. The flow of lubricant through the narrowing interfacial gap induces a hydrodynamic pressure change, which not only supports the externally applied load but can also induce significant deformation of the bounding surfaces. This deformation alters the local film thickness, creating a strongly non-linear feedback loop that must be accurately calculated for reliable predictions.

Furthermore, modelling lubricated interfaces becomes increasingly complex when real surface features that strongly influence interfacial behaviour are considered, such as microscale textures, stochastic roughness patterns, or compliant multilayered coatings. These surface features are engineered in many applications (\emph{e.g.}, biomechanical implants and automotive components) to tailor load support, friction, and durability. However, optimising such textures and features is inherently multifactorial, as performance depends on the combined effects of geometric design, operating conditions, lubricant rheology, and material properties, all of which dictate the interactions at the fluid-solid interfaces \cite{Sharma2021, Profito2024}. At the micro- and mesoscales, surface topography introduces localised geometric perturbations that alter the lubricant flow field, generating sudden pressure and temperature gradients, shear stress localisation, and highly non-uniform film thickness profiles across the contact interface \cite{Lu2020, Boidi2021, Vishnoi2021}, ultimately affecting the load-carrying capacity, friction and wear evolution of contacting systems. Further complicating this behaviour are thermally and pressure-driven phenomena, such as viscosity variations, phase transitions, and cavitation. Frictional heating can substantially modify the lubricant’s viscosity and its ability to maintain a stable fluid film. At the same time, the formation of pressure-induced cavitation pockets can disrupt flow continuity, reducing load support and accelerating surface damage \cite{Dowson1979Jan, Braun2010Sep}. Therefore, accurately representing these coupled multiphysics and multiscale processes poses a significant modelling challenge, as numerical approaches must simultaneously conserve transport quantities, remain stable in the presence of sharp gradients and non-linearities, and resolve advection–diffusion mechanisms with high geometric fidelity.

The robustness of the proposed EbFVM approach is evaluated by considering a set of lubrication problems pertaining to journal bearing systems as a representative engineering benchmark. Prevalent across rotating equipment and mechanical assemblies, journal bearings comprise a rotating shaft (or journal) housed within a stationary sleeve (or bushing), separated by a thin, dynamically evolving lubricating film (see \autoref{fig:Tribological_Interactions}). These systems pose significant numerical challenges for lubrication modelling, requiring high-resolution meshes to capture small deviations in shaft alignment caused by manufacturing imperfections, surface wear, vibration, or varying operational conditions, which can lead to asymmetries in pressure and temperature distributions, film thickness profiles, and fluid cavitation effects. Moreover, the cylindrical geometry adds complexity to the formulation and implementation of boundary conditions within the numerical solution.

\subsection{Governing Equations}
This section introduces the governing equations that describe the coupled fluid-solid interactions in lubricated journal bearings, accounting for two-dimensional hydrodynamic pressure generation, lubricant film thickness evolution due to shaft misalignments, and three-dimensional thermal dissipation induced by viscous shear. The framework adopted here is built on Reynolds-type approximations that link hydrodynamic and structural effects via robust multiphysical coupling models \cite{Khonsari2017Aug, Ardah2025}. A Cartesian coordinate system $\text{O}xyz$ is employed, with the $\text{O}x$-, $\text{O}y$-, and $\text{O}z$-axes corresponding to the circumferential, axial and film thickness directions, respectively.

\subsubsection{Generalised Reynolds Equation with Elrod-Adams \texorpdfstring{$(p$--$\theta)$ Cavitation Model}{Generalised Reynolds Equation with Elrod-Adams (p-theta) Cavitation Model}}
Given the tangential velocity fields of the journal $\boldsymbol{\mathrm{v}}_{j}(x,y,t) = \begin{bmatrix} u_1 & v_1 \end{bmatrix}^{T}$ and bushing $\boldsymbol{\mathrm{v}}_{b}(x,y,t) = \begin{bmatrix} u_2 & v_2 \end{bmatrix}^{T}$ surfaces in the $x$- and $y$-directions, toghether with the lubricant film thickness $h(x,y,t)$, density $\rho(x,y,t)$ and dynamic viscosity $\eta(x,y,t)$, the hydrodynamic pressure $p(x,y,t)$ and fluid film fraction $\theta(x,y,t)$ fields within the lubricant film are governed by the generalised Reynolds equation (GRE) \cite{Dowson1962Mar} in conjunction with the $(p$--$\theta)$ Elrod-Adams cavitation model \cite{Elrod1981}. The cavitation model introduces the auxiliary field $\theta$, which represents the fluid saturation in the cavitation regions, enabling a unified description of pressurised and cavitated regions. The $(p$--$\theta)$ GRE can be expressed as follows \cite{Dowson1962Mar}:
\begin{gather}
        {\underbrace{ \vphantom{\left(\frac{a^{0.3}}{b}\right)} \boldsymbol{\nabla} \cdot \Big( \boldsymbol{\Gamma}_{\mathrm{d}} \boldsymbol{\nabla} p \Big)}_{\mathclap{\substack{\text{Poiseuille} \\[0.5ex] \text{Term}}}}} = {\underbrace{ \vphantom{\left(\frac{a^{0.3}}{b}\right)} \boldsymbol{\nabla} \cdot \bigg[ \theta \Big( \boldsymbol{\Gamma}_{\mathrm{c}} \boldsymbol{\mathrm{v}}_{\mathrm{e}} + \boldsymbol{\Gamma}_{\mathrm{c1}} \boldsymbol{\mathrm{v}}_{\mathrm{1}} \Big) \bigg]}_{\mathclap{\substack{\text{Couette} \\[0.5ex] \text{Term}}}}} \; + \; {\underbrace{ \vphantom{\left(\frac{a^{0.3}}{b}\right)} \frac{\partial}{\partial t} \Big( \theta \rho_{e} \Big)}_{\mathclap{\substack{\text{Squeeze} \\[0.5ex] \text{Term}}}}},
        \label{eq:GRE}
\end{gather}
where $ \boldsymbol{\nabla}(\cdot) = \begin{bmatrix} \dfrac{\partial (\cdot)}{\partial x} & \dfrac{\partial (\cdot)}{\partial y} \end{bmatrix}^{T}$ is the gradient operator and $\boldsymbol{\mathrm{v}}_\mathrm{e} = \begin{bmatrix} \dfrac{u_1 + u_2}{2} & \dfrac{v_1 + v_2}{2} \end{bmatrix}^{T}$ is the average surface velocity vector. The tensors $\boldsymbol{\Gamma}_{\mathrm{d}}$, $\boldsymbol{\Gamma}_{\mathrm{c}}$, $\boldsymbol{\Gamma}_{\mathrm{c1}} \in \mathbb{R}^{2 \times 2}$ represent the effective diffusivity and convection coefficients, denoted as follows:
\begin{equation*}
        \boldsymbol{\Gamma}_{\mathrm{d}} = \varepsilon \textbf{I},
        \quad \quad \quad
        \boldsymbol{\Gamma}_{\mathrm{c}} = \rho^{*}_{e} \textbf{I},
        \quad \quad \quad
        \boldsymbol{\Gamma}_{\mathrm{c1}} = \rho^{*}_{1} \textbf{I},
\end{equation*}
where \textbf{I} is the identity tensor, and the auxiliary terms are expressed as follows:
\begin{align*}
        &\varepsilon  = \frac{\eta_e}{\eta^{'}_{e}} \rho^{'} - 
        \rho^{''},
        &\rho^{*}_{e} &= 2 \eta_{e} \rho^{'},
        \quad \quad 
        &\rho^{*}_{1} &= \rho_{e} - \rho^{*}_{e},
        \\[1ex]
        &\frac{1}{\eta_e} = \bigintsss_{z_1}^{z_2} \frac{dz}{\eta},
        &\frac{1}{\eta_{e}^{'}} &= \bigintsss_{z_1}^{z_2} \frac{z \; dz}{\eta},
        &\rho_{e}     &= \bigintsss_{z_1}^{z_2} \rho \; dz,
\end{align*}
\begin{align*}
        \rho^{'} = \bigintsss_{z_1}^{z_2} \rho \; \left( \bigintsss_{z_1}^{z} \frac{dz'}{\eta} \right) \; dz,
        \quad \quad \quad \quad
        \rho^{''} = \bigintsss_{z_1}^{z_2} \rho \; \left( \bigintsss_{z_1}^{z} \frac{z'dz'}{\eta} \right) \; dz.
\end{align*}

Assuming that the cavitated regions are composed of a homogeneous mixture of liquid and gas/vapour phases, and adopting the rule of mixtures for density estimation, the complementarity boundary conditions for cavitation are given as follows:
\begin{align}
        (p - p_\text{cav})(1 - {\theta}) = 0 \; {\rightarrow} \; 
        \begin{cases}
           p > p_\text{cav} \quad {\rightarrow} \quad {\theta}=1 & \text{in {$\mathfrak{D}^+$}},
           \\[1ex]
           p = p_\text{cav} \quad {\rightarrow} \quad 0\;{\leq}\;{\theta}<1 &    \text{in {$\mathfrak{D}^0$}},
        \end{cases}
        \label{eq:p-theta}
\end{align}
where $p_{\text{cav}}$ is the limit cavitation pressure, $\mathfrak{D}^+$ is the pressured regions within the lubricated domain, while $0\;{\leq}\;{\theta}<1$ indicates the breakdown of the lubricant film inside the cavitated regions $\mathfrak{D}^0$, signifying the presence of a homogeneous mixture of liquid and gases/vapours in those regions. A fully developed lubricant film is achieved when $\theta=1$. Detailed implementations of the Elrod–Adams formulation, including its integration in the GRE model, are provided in \cite{Ausas2009Jul, Profito2015Sep, Ardah2023Jan}, as well as algorithmically summarised in \ref{sec:Appendix_p-theta}.

\subsubsection{Thermal Energy Equation}
The thermal behaviour of the lubricant film, influenced by both friction-induced heating and viscous dissipation, is governed by the energy conservation equation for a compressible, viscous medium. In conservative form, the temperature field $T(x,y,z,t)$ is described as follows:
\begin{gather}
        \frac{\partial}{\partial t} \Big( \rho c_{p} T \Big)
        + \boldsymbol{\nabla} \cdot \Big( \rho c_{p}  \boldsymbol{v} T \Big) = \boldsymbol{\nabla} \cdot \Big( k \boldsymbol{\nabla} T \Big) + Q_{T},
        \label{eq:FilmEnergy_Cnsrv}
\end{gather}
where $\rho,$ $c_p$ and $k$ denote the local density, specific heat capacity and thermal conductivity of the lubricant, respectively. The source term $Q_T$ encapsulates all volumetric heat generation sources expressed as follows:
\begin{gather*}
        Q_{T} = \Big( Q_{p} + Q_{cp} + Q_{\Phi} + \dot q_{v} \Big),
        \\[1.5ex]
        Q_{p} = \underbrace{{\beta} T \left( \frac{\partial p}{\partial t} + \boldsymbol{v} \cdot \boldsymbol{\nabla}p \right)}_{\substack{\text{Compressive} \\ \text{Heating/Cooling}}},
        \quad \quad \quad 
        Q_{cp} = \underbrace{{\rho} T \left( \frac{\partial c_{p}}{\partial t} + \boldsymbol{v} \cdot \boldsymbol{\nabla} c_{p} \right)}_{\substack{\text{Enthalpic} \\ \text{Heating/Cooling}}}, \nonumber
        \quad \quad  \quad 
        Q_{\Phi} = \underbrace{\eta \left[ \left( \frac{\partial u}{\partial z}\right)^2 + \left( \frac{\partial v}{\partial z}\right)^2 \right]}_{\substack{\text{Viscous} \\ \text{Heating}}}, \nonumber
\end{gather*}
where $\beta$ is the thermal compressibility of the lubricant and $\boldsymbol{v}(x,y,z,t) = u\,\hat{\mathbf{i}} + v\,\hat{\mathbf{j}} + w\,\hat{\mathbf{k}}$ is the velocity vector field of the lubricant film. The internal heat source $\dot{q}_v$ accounts for dissipation mechanisms such as asperity interactions and chemical reactions; however, these effects are neglected in the present study for simplicity.

\subsubsection{Lubricant Film Thickness}
\begin{figure}[H]
   \centering
   \includegraphics[width=\linewidth]{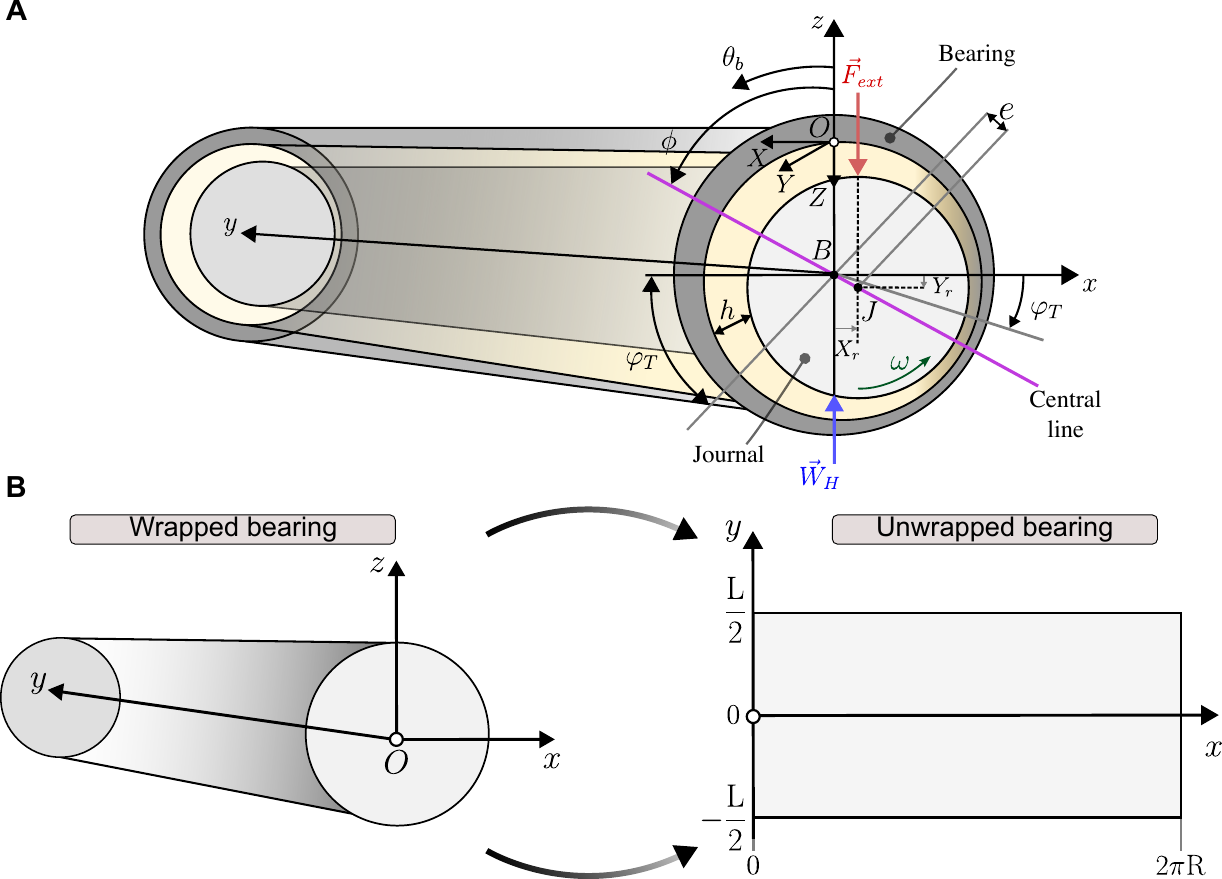}
   \caption{Schematic of the simulated journal bearing system. (\textbf{A}) Cross-sectional view highlighting key geometric features and nomenclature. (\textbf{B}) Transformation of the cylindrical geometry into an unwrapped two-dimensional domain, justified by the negligible influence of curvature on hydrodynamic behaviour.}
   \label{fig:JB_CoordinateSystem}
\end{figure}

Under the assumption of rigid contacting surfaces, the spatiotemportal variation of the lubricant film thickness $h(x, y, t)$ is expressed in the local coordinate system $Oxyz$ as:
\begin{gather}
        h(x, y, t) = c - (Y_r - A_r y) \cos (x) + (X_r - B_r y) \sin (x) + h_T(x, y, t),
\end{gather}
where $c$ denotes the nominal radial clearance between the journal and bushing, and $h_T(x, y)$ captures the contribution from surface texture or microscale topographical features. The parameters ($X_r$, $Y_r$) and ($A_r$, $B_r$) define the lateral displacements and angular tilts of the journal centerline with respect to the bearing axis, respectively, thereby accounting for shaft misalignment effects, as illustrated in \autoref{fig:JB_CoordinateSystem}\textbf{A}. 

To resolve the non-linearities inherent in the film thickness solution, particularly under conditions of dynamic misalignment and complex geometrical variations, a Newton–Raphson iterative scheme is adopted. This solver is further stabilised using an Armijo line search algorithm, which adaptively controls the step size to ensure robust global convergence. The numerical strategy follows the methodology outlined in \cite{Profito2015}, focusing solely on fluid-mediated load support and excluding direct surface contact or solid-solid friction effects. A detailed mathematical formulation and implementation procedure for the iterative scheme are provided in \ref{sec:Appendix_Misalignment}.

\subsubsection{Thermophysical Characterisation of Lubricant}
The density and dynamic viscosity of the lubricant are modelled as functions of local pressure and temperature to account for compressibility and thermal effects. The density $\rho$ is updated according to the Dowson and Higginson pressure–temperature–viscosity relationship expressed as \cite{Hamrock2004Mar}:
\begin{equation}
        \rho(p, T, t) = \rho_0 \Bigg( 1 + \frac{C_1 p}{1 + C_2 p} \Bigg) \Bigg[ 1 - C_3 (T - T_0) \Bigg],
\end{equation}
where $\rho_0$ is the density at the ambient temperature $T_0$, while $C_1$, $C_2$ and $C_3$ are empirically determined coefficients that describe the compressibility and thermal expansion of the fluid, given as $C_1 = 0.6 \times 10^9 \; \mathrm{Pa}$, $C_2 = 1.7 \times 10^9 \; \mathrm{Pa}$ and $C_3 = 6.5 \times 10^{-4} \; \mathrm{K^{-1}}$.

The dynamic viscosity $\eta$ incorporating both pressure-thickening and temperature-thinning effects is expressed via the Roelands pressure–temperature–viscosity relationship written as \cite{Roelands1966}:
\begin{equation}
        \eta(p, T, t) = {\eta_0} \times \text{exp} \left\{ \Big[ {\ln}({\eta_0}) + 9.67 \Big] \left( -1 + {\left(1 + \frac {p}{p_{r0}}\right)}^{Z} \times \left( {\frac{T - 138}{{T_0} - 138}} \right)^{-S_0}  \right) \right\}
\end{equation}
where $\eta_0$ is the reference viscosity, $T_0$ is the ambient temperature (in Kelvin), $p_{r0} = 5.1 \times 10^9 \;  \mathrm{Pa}$ is the reference pressure, and $Z$, $S_0$ and $\beta^*$ are fitted parameters that account for pressure–viscosity and thermal–viscosity sensitivities, given as $Z = 0.689$ and $S_0 = 1.3891$.

Under cavitation, the lubricant film may exist as a two-phase mixture of liquid and vapour. To model the resulting spatiotemporal variation in lubricant properties, the effective rheological quantities are updated using a linear mixture formulation weighted by the local liquid fraction $\theta \in [0,1]$ computed via \autoref{eq:GRE}. Therefore the rupture-induced lubricant properties are adjusted as follows:
\begin{equation}
    \begin{aligned}
            \eta(p, T, \theta, t) &\rightarrow \theta \cdot \eta(p, T, t), \\[1ex]
            \rho(p, T, \theta, t) &\rightarrow \theta \cdot \rho_(p, T, t).
    \end{aligned}
\end{equation}

\subsection{Simulation Setup}
\label{sec:SimulationSetup}
The computational grids employed in the upcoming simulations are all constructed in \textsc{Abaqus} \cite{Abaqus2012} to maintain precise geometry resolution and accommodate adaptive meshing. The inherently cylindrical geometry of the journal bearing is reformulated as an \emph{unwrapped} two-dimensional rectangular domain for numerical efficiency, as shown in \autoref{fig:JB_CoordinateSystem}\textbf{B}. This transformation is justified by the small curvature ratio typically observed in journal bearing configurations, whereby $\frac{c}{R} \ll 1$, where $c$ is the radial clearance and $R$ is the shaft radius \cite{Szeri2010Dec}. Under this approximation, any curvature effects incurred on to the hydrodynamic performance are neglected, hence permitting the use of a local Cartesian coordinate system to represent the bearing surface. Unless stated otherwise, all simulations utilise the geometric, material and operating parameters summarised in \autoref{tab:SimulationParameters}.

As for the boundary conditions, periodic constraints are applied along the circumferential (or vertical) ends of the unwrapped bearing domain to ensure mass continuity of the flowing lubricant, while ambient pressure and temperature conditions are applied along the axial (or horizontal) ends of the rectangular bearing domain, as illustrated in the computational domain setup in \autoref{fig:JB_CoordinateSystem}\textbf{B}. When a lubricant supply hole (feeding hole) is present at the bearing inlet, the mesh nodes located within the hole are assigned the same pressure and temperature values as those of the supply lubricant. This enables accurate depiction of the boundary conditions at the supply entry, where the lubricant enters the bearing at predefined environmental conditions (see \autoref{tab:SimulationParameters}).

\begin{table}[H]
    \centering
    \renewcommand{\arraystretch}{1.2}
    \setlength{\tabcolsep}{11pt}
    \caption{Simulation parameters utilised in the simulation of the journal bearing system.}
    \label{tab:SimulationParameters}
    \begin{tabular}{p{4.5cm} p{6.5cm} p{4cm}}
        \hline
         & \textbf{Parameter} & \textbf{Value} \\
        \hline
        \multirow{5}{*}{\textbf{Geometry}} 
            & Bearing radius, $R_b$ & $\SI{30}{mm}$ \\
            & Bearing width, $L$ & $\SI{80}{mm}$ \\
            & Radial clearance, $c$ & $\SI{20}{\micro\metre}$ \\
            & Feeding hole radius, $R_h$ & $\SI{6}{mm}$ \\
            & Features depth, $h_T$ & $\SI{20}{\micro\metre}$ \\
        \hline
        \multirow{3}{*}{\textbf{Operating Conditions}} 
            & Rotational speed, ${[u_1 \; v_1]}$ & [5000 0] rev/min \\
            & External force, ${[F_{x} \; F_{y}]}$ & [0 -8000] N \\
            & External moment, ${[W_{x} \; W_{y}]}$ & [0 800] Nm \\
        \hline
        \multirow{5}{*}{\textbf{Lubricant Properties}} 
            & Ambient density, $\rho_0$ & $\SI{810}{kg/m^3}$ \\
            & Ambient viscosity, $\mu_0$ & $\SI{0.1}{Pa \; s}$ \\
            & Thermal conductivity, $k$ & $\SI{0.105}{W/(m \; K)}$ \\
            & Thermal capacity, $c_p$ & $\SI{2300}{J/(kg \; K)}$ \\
            & Supply temperature, $T_s$ & $\SI{80}{^\circ C}$ \\
        \hline
        \multirow{3}{*}{\textbf{Boundary Conditions}} 
            & Ambient pressure, $p_{\text{amb}}$ & $\SI{100}{kPa}$ \\
            & Ambient temperature, $T_{\text{amb}}$ & $\SI{80}{^\circ C}$ \\
            & Cavitation pressure, $p_{\text{cav}}$ & $\SI{0}{Pa}$ \\
        \hline
    \end{tabular}
\end{table}

\section{Results}
\label{sec:Results}
In the sections that follow, we present a series of comparative results designed to evaluate the performance and versatility of the EbFVM across element types and interface configurations. All simulations are based on the input parameters and boundary conditions outlined in \autoref{sec:SimulationSetup} to ensure a consistent basis for comparison. Performance metrics include two-dimensional spatial distributions of the hydrodynamic pressure, lubricant film fraction, lubricant thickness and film temperature distribtions. The temperature fields presented herein are extracted along the mid-plane ($xy$-plane) of the lubricant film, capturing the thermal behaviour at its central layer.

\subsection{Standard Journal Bearing Configuration}
We begin with a controlled benchmark comparison using a simple cylindrical journal bearing configuration to evaluate the accuracy and robustness of the EbFVM across three numerical strategies: (i) EbFVM using hexahedral elements, (ii) EbFVM using triangular prism elements and (iii) classical structured finite volume method (FVM) implemented on a regular hexahedral mesh. Readers are referred to \cite{Ardah2023Oct} for further implementation details of the vertex-centered finite volume model, which will serve as the baseline in the subsequent validations.

\begin{figure}[hb!]
   \centering
   \includegraphics[width=\linewidth]{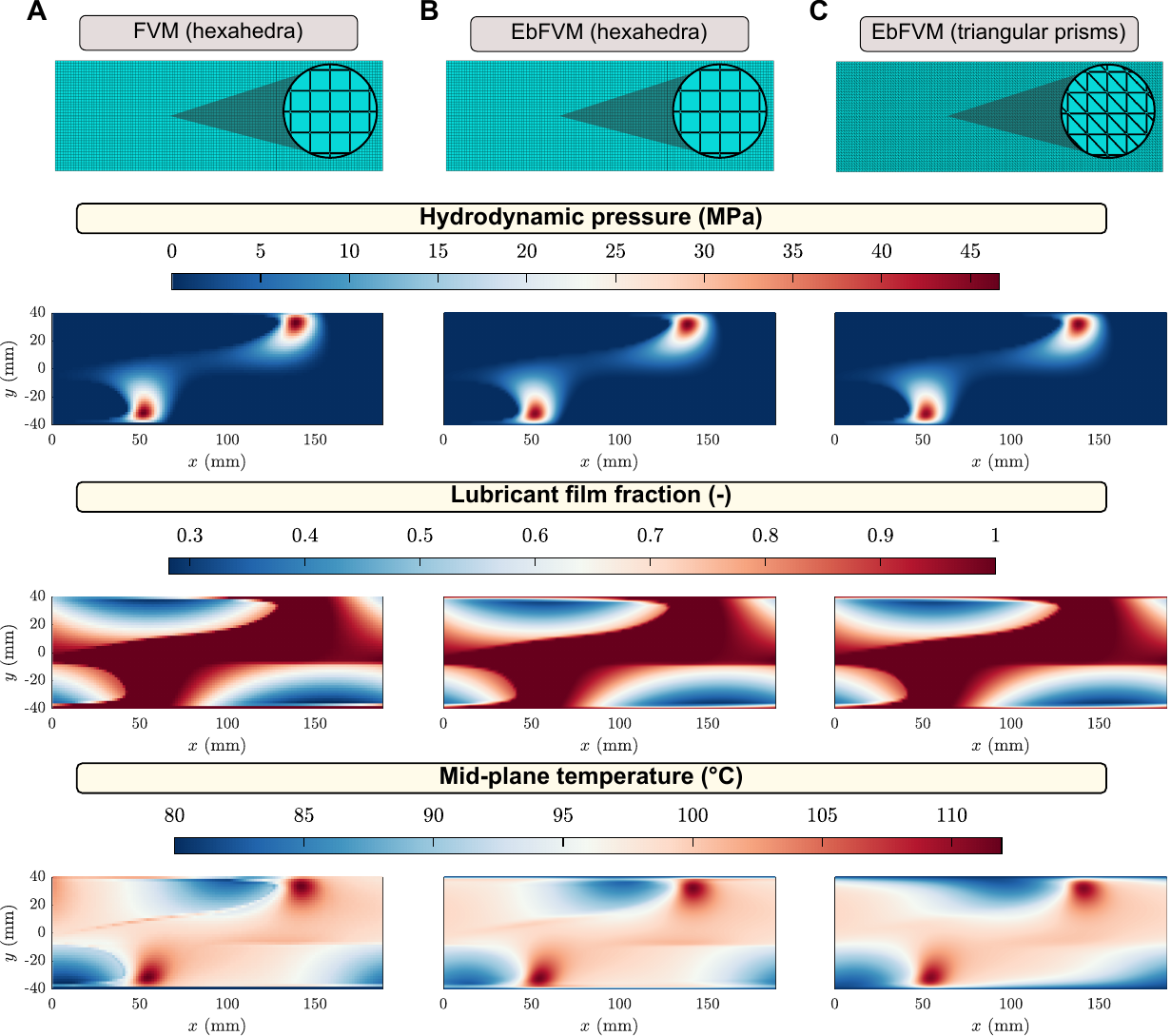}
   \caption{Comparative evaluation of numerical discretisation schemes based on key lubrication field predictions, comprising hydrodynamic pressure, lubricant fraction and mid-plane temperature computed using (\textbf{A}) conventional hexahedral finite volume method,  (\textbf{B}) hexahedral element-based finite volume method and (\textbf{C}) prismatic element-based finite volume method.}
   \label{fig:JBR_EbFVM_FVM}
\end{figure}

As illustrated in \autoref{fig:JBR_EbFVM_FVM}, the EbFVM approach, implemented with either hexahedral or triangular prism elements, yields results that are in excellent agreement with the structured FVM solutions, particularly with respect to hydrodynamic pressure and lubricant film fraction. The impact of the periodic boundary conditions applied at the circumferential ends of the computational domain is clearly reflected across all simulations, as evidenced by the repeating features and symmetry in the pressure and cavitation fields, confirming their consistent enforcement across all discretisation strategies. The observed pressure asymmetry arises from shaft misalignment along both the axial and longitudinal directions, a key geometric perturbation that all three numerical strategies are able to capture. Consequently, the predicted lubricant fraction fields exhibit strong agreement, indicating that the EbFVM is capable of resolving cavitation due to its inherent mass-conservation properties.

Despite the overall strong agreement in the predicted thermo-hydrodynamic fields, localised discrepencies are observed near the axial boundaries of the bearing (\emph{i.e.}, around $y = \pm 40 \, \text{mm}$), particularly in the thermal field predicted by the prismatic-element EbFVM. While both the conventional structured FVM formulation and the hexahedral-element EbFVM yield nearly identical temperature distributions across the entire domain, the prismatic-element EbFVM predicts slightly lower temperatures near the horizontal boundaries. The largest deviations occur near $x = 50 \, \text{mm}$ at the upper wall ($y = + \,40 \, \text{mm}$) and $x = 150 \, \text{mm}$ at the lower wall ($y = - \,40 \, \text{mm}$), where the mean absolute error reaches approximately 1.02 $\mathrm{^oC}$ relative to the hexahedral-element solutions. Nonetheless, the domain-averaged temperature discrepancy remains below 4$\%$, confirming that the global energy balance is well preserved. 

\begin{figure}[H]
   \centering
   \includegraphics[width=\linewidth]{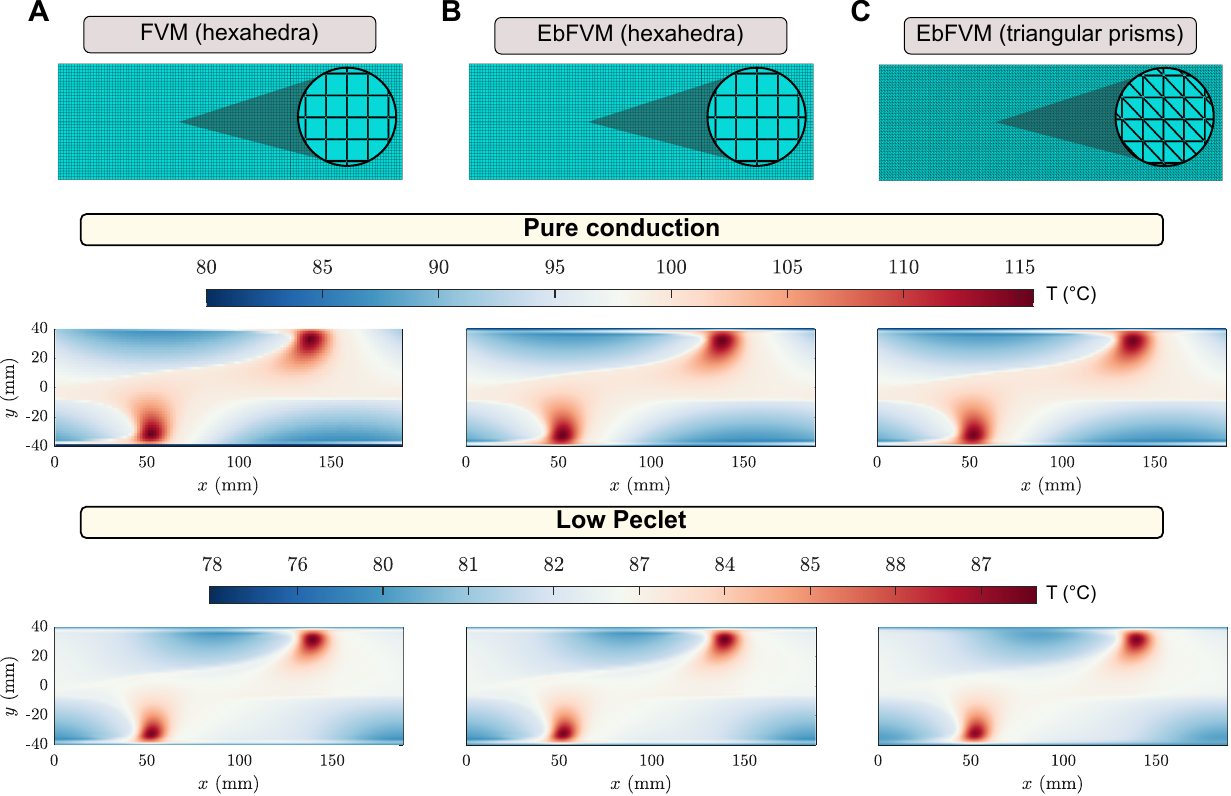}
   \caption{Verification of the element-based finite volume method (EbFVM) under conduction-dominated and low Peclet regimes. Comparison of temperature distributions predicted using (\textbf{A}) the conventional hexahedral finite volume method, (\textbf{B}) the hexahedral EbFVM, and (\textbf{C}) the prismatic EbFVM for pure conduction and convection–diffusion cases at low Peclet number. All models exhibit nearly identical temperature fields, confirming numerical consistency of the EbFVM formulations across element types under the aforementioned operating conditions.}
   \label{fig:JBR_EbFVM_FVM_Conduction_LowPeclet}
\end{figure}

As all simulations exhibit the same physical boundary conditions and spatial resolution, the observed differences cannot be attributed to mesh density or geometry. Hence, it can be inferred that discrepancies in the thermal solution are caused by the element-type–dependent discretisation characteristics, particularly in the evaluation of convective fluxes within the prismatic control volumes. The prismatic EbFVM introduces slight geometric asymmetries in the sub-control-volume layout, as shown in \autoref{fig:JBR_EbFVM_FVM}\textbf{A}, which can alter the local balance between axial advection and radial heat conduction, especially under steep temperature gradients close to the axial ends. This interpretation is supported by verification tests under pure conduction and low Peclet number conditions demonstrated in \autoref{fig:JBR_EbFVM_FVM_Conduction_LowPeclet}, for which the prismatic and hexahedral EbFVM (and FVM) formulations yield matching temperature (and hydrodynamic fields). The discrepancies therefore emerge only at higher Peclet regimes where convective transport dominates. Such behaviour is consistent with known numerical sensitivities of upwind-biased convection schemes on non-orthogonal prism elements and does not indicate a loss of conservation or stability. While further analysis or experimental validation would be needed to identify which prediction best represents the physical temperature field, the quantified discrepancies remain within expected limits for convection-dominated thermo-hydrodynamic simulations. Overall, the results confirm that the EbFVM maintains high accuracy and conservation across element types, with the prismatic formulation successfully reproducing all key thermo-hydrodynamic trends despite insignificant geometric sensitivity near the axial walls.

\subsection{Micro-Textured Lubricated Interfaces}
\label{subsec:MicroTexture}
Micro-texturing is an effective surface engineering strategy for optimising tribological performances of lubricated mechanical systems \cite{Gropper2016, Grützmacher2019}. Micro-textures at the microscale enhance hydrodynamic performance by inducing local pressure variations, enhancing lubricant flow and controlling shear stress levels, leading to improved lift, reduced friction and delayed film rupture. Such textures are often engineered using advanced fabrication techniques such as laser surface texturing \cite{Etsion2005}, photoetching etching \cite{Hamilton1966} and precision grinding \cite{Silva2017}, allowing for controlled replication of dimples, grooves or chevron-like structures across the contact surface. In the context of journal bearings, micro-texturing provides a means to mitigate some of the key challenges encountered in high-speed, high-load environments, particularly those related to lubricant starvation, frictional heating and vibration-induced wear \cite{Profito2024}.

In the field of lubrication modelling, micro-textured bearings are commonly analysed using commercial and open-source Computational Fluid Dynamics (CFD) solvers such as OpenFOAM, COMSOL Multiphysics and Ansys Fluent \cite{Cupillard2008, Wang2018, Marian2022}. These simulation platforms provide extensive flexibility for resolving complex transport phenomena across intricate interfacial geometries. However, their application can become computationally prohibitive when deployed to tackle thin-film lubrication problems. In addition, capturing cavitation and strongly coupled thermo-viscous effects within unstructured domains necessitates manual mesh tuning, advanced stabilisation schemes and rigorous numerical calibration, thereby limiting their practicality and scalability for lubrication analyses. In contrast, the EbFVM developed in this study offers a tailored framework for modelling interfacial transport processes. By rigorously enforcing local conservation and enabling seamless geometric adaptability across polyhedral elements, the EbFVM achieves robust flux evaluation even in the presence of sharp surface discontinuities. Its element-centred formulation permits local refinement around geometrical intrusions (\emph{e.g.}, dimples, grooves, feeding holes) without requiring re-meshing or interpolation. Moreover, its significantly reduced computational overhead relative to traditional CFD packages enables fast and scalable simulations, making it well-suited for rapid performance benchmarking of complex interfacial systems.

Therefore, this section corroborates the strengths of the EbFVM by extending the analysis to journal bearings systems with engineered surface textures and embedded lubrication supply holes, features that require localised mesh refinement to accurately capture the flow patterns. For these geometrically complex configurations, hexahedral elements are employed within the EbFVM framework, owing to the higher number of integration points compared to prismatic elements for improved accuracy of flux evaluation over intricately highly textured surfaces and multiphysics interactions.

\begin{figure}[H]
   \centering
   \includegraphics[width=\linewidth]{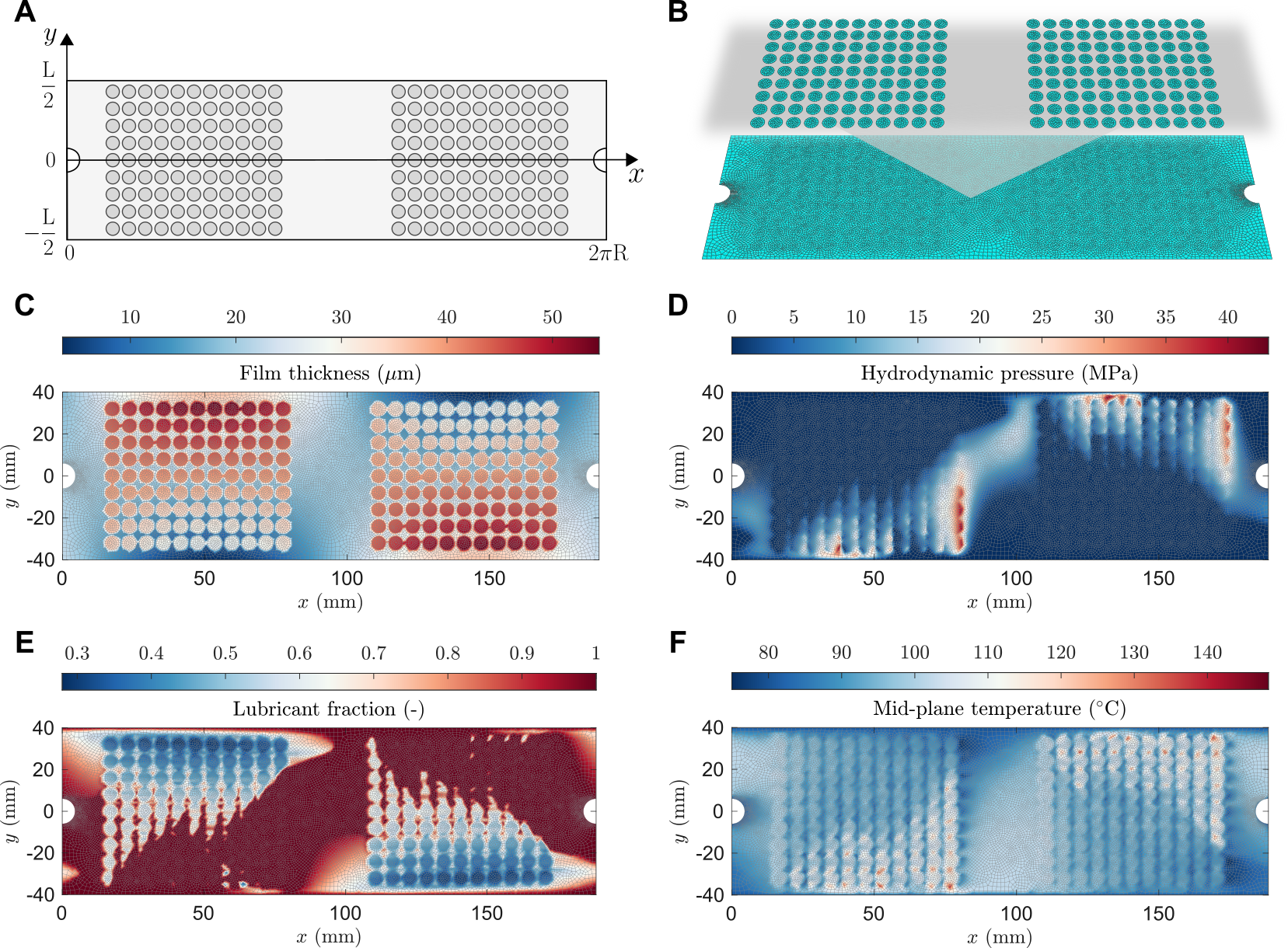}
   \caption{Simulated thermo-hydrodynamic behaviour of a textured journal bearing incroporating circular surface dimples and an inlet feeding hole. (\textbf{A}) Unwrapped view of the bearing surface featuring circumferentially distributed circular dimples. (\textbf{B}) Computational mesh illustrating targeted refinement in the periphery of dimples and the supply hole. (\textbf{C}) Film thickness profile showing topographical modulation from dimple texture. (\textbf{D}) Hydrodynamic pressure distribution capturing micro-hydrodynamic effects and flow-induced loading. (\textbf{E}) Lubricant fraction map indicating pressure-driven cavitation regions and reformation zones. (\textbf{F}) Mid-plane temperature distribution revelaing the influence of frictional heating and convective–conductive heat transport on the local thermal gradients.}   
   \label{fig:JBR_Dimples}
\end{figure}

We first consider a journal bearing configuration incorporating periodically distributed circular surface dimples designed to enhance lubrication performance by altering local flow dynamics and pressure generation mechanisms. These dimples serve as micro-reservoirs that promote lubricant entrainment and reducing shear-induced energy losses, especially under full-film lubrication conditions such as in automotive internal combustion engines \cite{Ronen2001}. Their geometric influence on the boundary layer flow gives rise to micro-hydrodynamic pressure pockets, which locally augment load-carrying capacity and contribute to the stabilisation of the lubricant film across the contact interface \cite{Vlădescu2019, Profito2024}. As shown in \autoref{fig:JBR_Dimples}\textbf{A}, the dimples, each with a depth of $h_T = \SI{20}{\micro\metre}$, are uniformly arranged along the circumferential direction, and a spatially refined mesh is employed to resolve both the textured surface features and the embedded supply inlet hole as illustrated in \autoref{fig:JBR_Dimples}\textbf{B}. 

The resulting film thickness distribution (\autoref{fig:JBR_Dimples}\textbf{C}) showcases localised elevation over the textured regions, while the pressure distribution (\autoref{fig:JBR_Dimples}\textbf{D}) exhibits local build-up that are characteristic of micro-hydrodynamic enhancement. The lubricant fraction map (\autoref{fig:JBR_Dimples}\textbf{E}) captures partial cavitation trailing the dimple regions, predominantly in regions where the film thickness reaches local maxima and the hydrodynamic pressure drops to its lowest values. Finally, the mid-plane temperature distribution (\autoref{fig:JBR_Dimples}\textbf{F}) reveals smooth thermal gradients without abrupt transitions, with peak temperatures confined to small, localised regions where cavitation is absent (\emph{i.e.} $\theta \rightarrow 1$) and viscous dissipation is most pronounced. When compared with the non-textured configuration demonstrated in \autoref{fig:JBR_NoTexture} (see \ref{sec:Appendix_JBR_NoTexture}), which serves as a baseline for evaluating the effect of surface patterning, the presence of dimples leads to a notably different flow and pressure behaviour. In the smooth bearing interface, the film thickness remains relatively uniform along the circumferential direction, producing a continuous pressure zone without the discrete peaks characteristic of the textured surface. Consequently, the lubricant fraction field in the non-textured case exhibits a more gradual transition between fully flooded and cavitated regions. The comparison between the two interfaces highlights the hydrodynamic and thermal effects of micro-dimple texturing.

\begin{figure}[H]
   \centering
   \includegraphics[width=\linewidth]{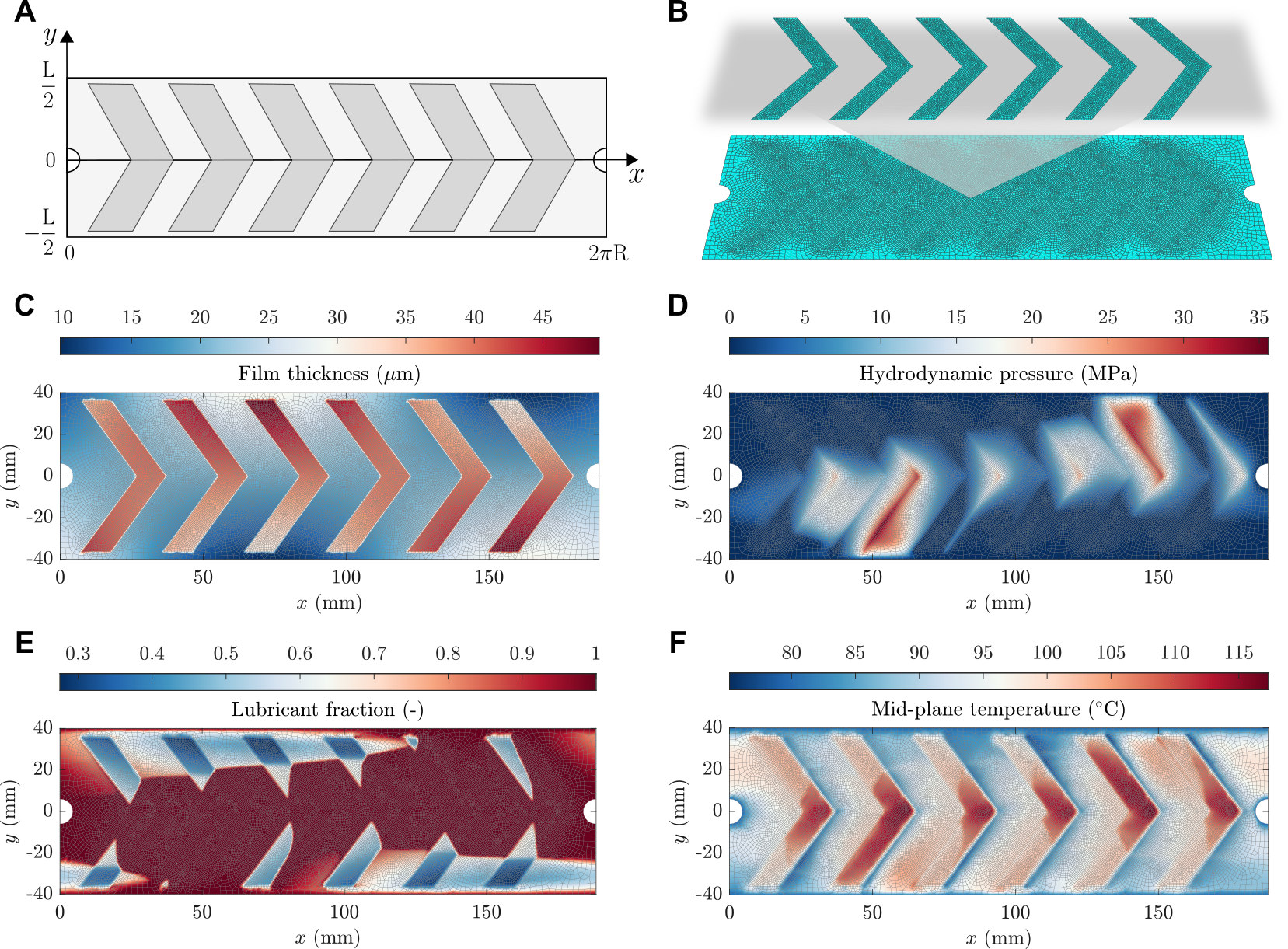}
   \caption{Simulated thermo-hydrodynamic behaviour of a textured journal bearing incroporating herringbone grooves and an inlet feeding hole. (\textbf{A}) Unwrapped view of the bearing surface featuring circumferentially distributed chevron-shaped grooves. (\textbf{B}) Computational mesh illustrating targeted refinement in the periphery of grooves and the supply hole. (\textbf{C}) Lubricant film thickness profile showing alternating thick and thin regions aligned with groove geometry. (\textbf{D}) Hydrodynamic pressure distribution capturing micro-hydrodynamic effects and flow-induced loading. (\textbf{E}) Lubricant fraction map illustrating cavitation downstream of groove flanks. (\textbf{F}) Mid-plane temperature field highlighting thermo-hydrodynamic effects induced by groove-driven flow and frictional heating.}   
   \label{fig:JBR_Herringbone}
\end{figure}

In contrast to the aforementioned micro-dimple features, \autoref{fig:JBR_Herringbone} examines the use of several herringbone grooves, which are elongated, V-shaped surface features aimed at imposing directional control over the lubricant flow and promoting hydrodynamic recovery within the converging-diverging interface \cite{Costa2007}, as has been demonstrated in turbochargers \cite{Liu2023} and water-based lubricated systems \cite{Xu2025}. The geometric domain, shown in \autoref{fig:JBR_Herringbone}\textbf{A}, comprises six herringbone grooves evenly distributed along the circumference, each with a depth of $h_T = \SI{20}{\micro\metre}$. This pattern is correspondingly captured in the lubricant film thickness distribution shown in \autoref{fig:JBR_Herringbone}\textbf{C}. As previously carried out for circular dimple setting, the computational domain is discretised using a refined unstructured mesh (\autoref{fig:JBR_Herringbone}\textbf{B}) to ensure accurate resolution of the grooves and lubricant supply inlet. The hydrodynamic pressure field (\autoref{fig:JBR_Herringbone}\textbf{D}) shows higher magnitudes along the groove convergence zones, particularly near $x = \SI{50}{\milli\metre}$ and $x = \SI{150}{\milli\metre}$. This behaviour directly correlates with the lubricant fraction map (\autoref{fig:JBR_Herringbone}\textbf{E}), where fully flooded regions align with zones of maximum pressure, and partial cavitation trails the diverging flanks where pressures decay. Accordinlgy, the mid-plane temperature field (\autoref{fig:JBR_Herringbone}\textbf{F}) shows temperature peaks ($\approx$ 110°C - 117°C) occurring near the tip regions of the chevron patterns characterised by minimal cavitation. The lower peak temperature, compared with that observed for the circular dimple configuration ($\approx$ 145 °C), highlights the strong influence of surface topography on the thermo-hydrodynamic response of the lubricated interface, even under identical operating conditions.

\begin{figure}[H]
   \centering
   \includegraphics[width=\linewidth]{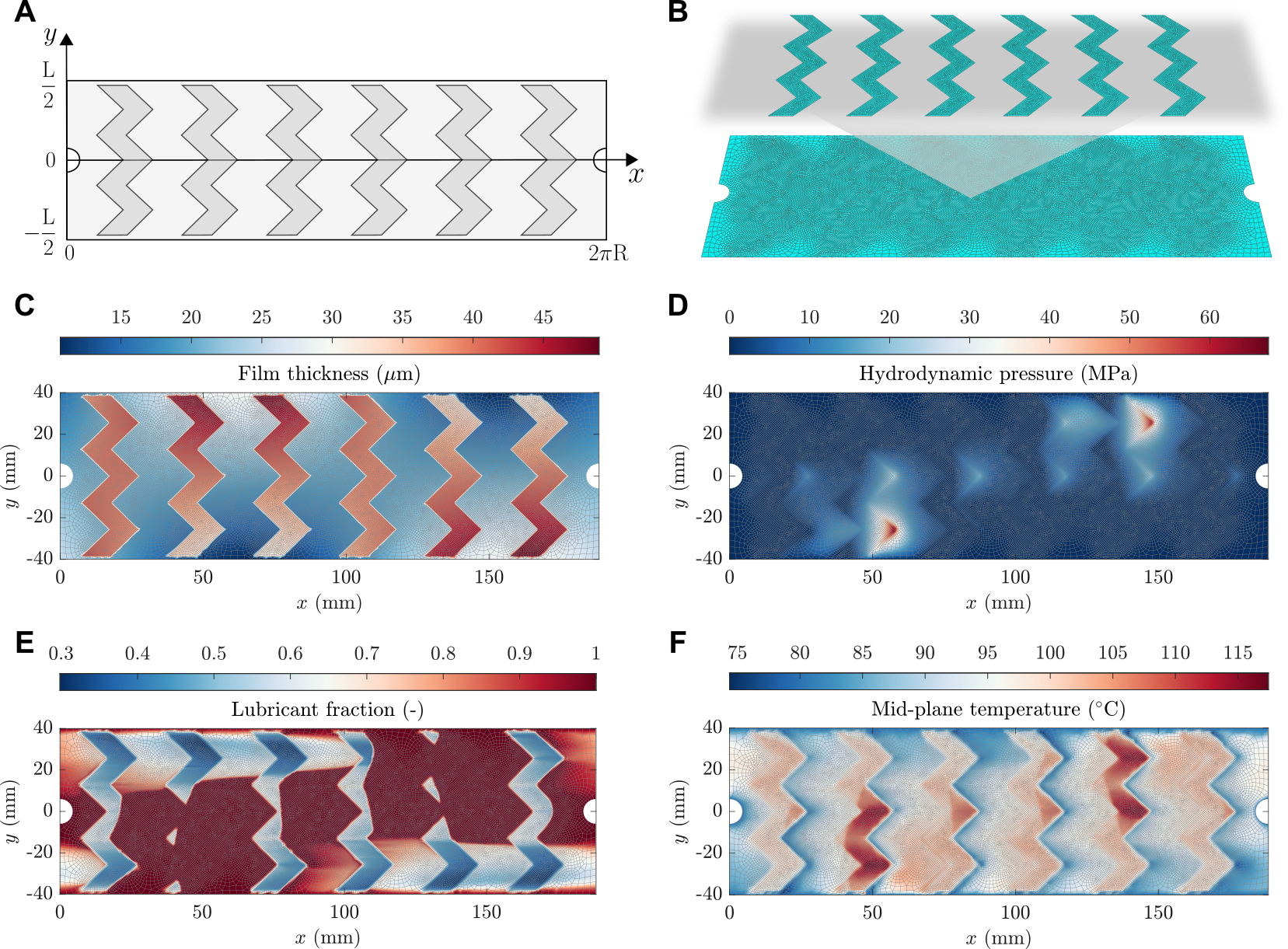}
   \caption{Simulated lubrication characteristics of a textured journal bearing with sawtooth grooves. (\textbf{A}) Layout of the unwrapped bearing surface with circumferentially distributed sawtooth-shaped grooves. (\textbf{B}) Computational mesh with targeted refinement around groove features and inlet supply hole. (\textbf{C}) Film thickness profile showing alternating thick and thin regions aligned with groove geometry. (\textbf{D}) Hydrodynamic pressure revealing local pressure build-up at groove converging zones. (\textbf{E}) Lubricant fraction map illustrating cavitation downstream of groove flanks. (\textbf{F}) Mid-plane temperature highlighting thermo-hydrodynamic effects induced by groove-driven flow and frictional heating.}   
   \label{fig:JBR_SawTooth}
\end{figure}

The journal bearing configuration presented in \autoref{fig:JBR_SawTooth} incorporates circumferentially aligned sawtooth grooves composed of three mini-chevron segments, each with a depth of $h_T = \SI{20}{\micro\metre}$. This texture is designed to promote asymmetric flow and directional pressure redistribution along the lubricated interface \cite{Filgueira2021}. The unwrapped surface representation (\autoref{fig:JBR_SawTooth}\textbf{A}) and the refined computational mesh (\autoref{fig:JBR_SawTooth}\textbf{B}) capture the pronounced geometric gradients induced by the sawtooth grooves and the inlet supply region. The resulting film thickness field (\autoref{fig:JBR_SawTooth}\textbf{C}) reflects these geometric variations, producing localised hydrodynamic pressure peaks at the leading edges of the grooves, particularly near \(x = \SI{50}{\milli\metre}\) and \(x = \SI{150}{\milli\metre}\) (\autoref{fig:JBR_SawTooth}\textbf{D}). These pressure maxima are followed by sharp declines along the diverging groove paths, illustrating the asymmetric flow behaviour generated by the sawtooth geometry. Furthermore, the lubricant fraction distribution (\autoref{fig:JBR_SawTooth}\textbf{E}) reveals that cavitation predominantly develops along the groove channels, while highly pressurised regions remain fully flooded. The mid-plane temperature field (\autoref{fig:JBR_SawTooth}\textbf{F}) exhibits smooth yet spatially extended gradients, with the highest film temperatures occurring within the full-film zones. These numerical results are consistent with experimental findings by Filgueira et al. \cite{Filgueira2021}, who demonstrated that directional surface texturing can effectively tailor lubrication performances. The present simulations extend that understanding by providing detailed mechanistic insight into how asymmetric shear, localized recirculation and groove-induced cavitation interact to influence pressure and temperature fields. Such predictive modelling offers a valuable tool for the optimisation of groove geometry and orientation, enabling the design of textured journal bearings with enhanced load-carrying capacity, controlled cavitation and improved thermal performance.

\section{Concluding Remarks and Perspectives}
\label{sec:Conclusion}
\begin{figure}[H]
   \centering
   \includegraphics[width=\linewidth]{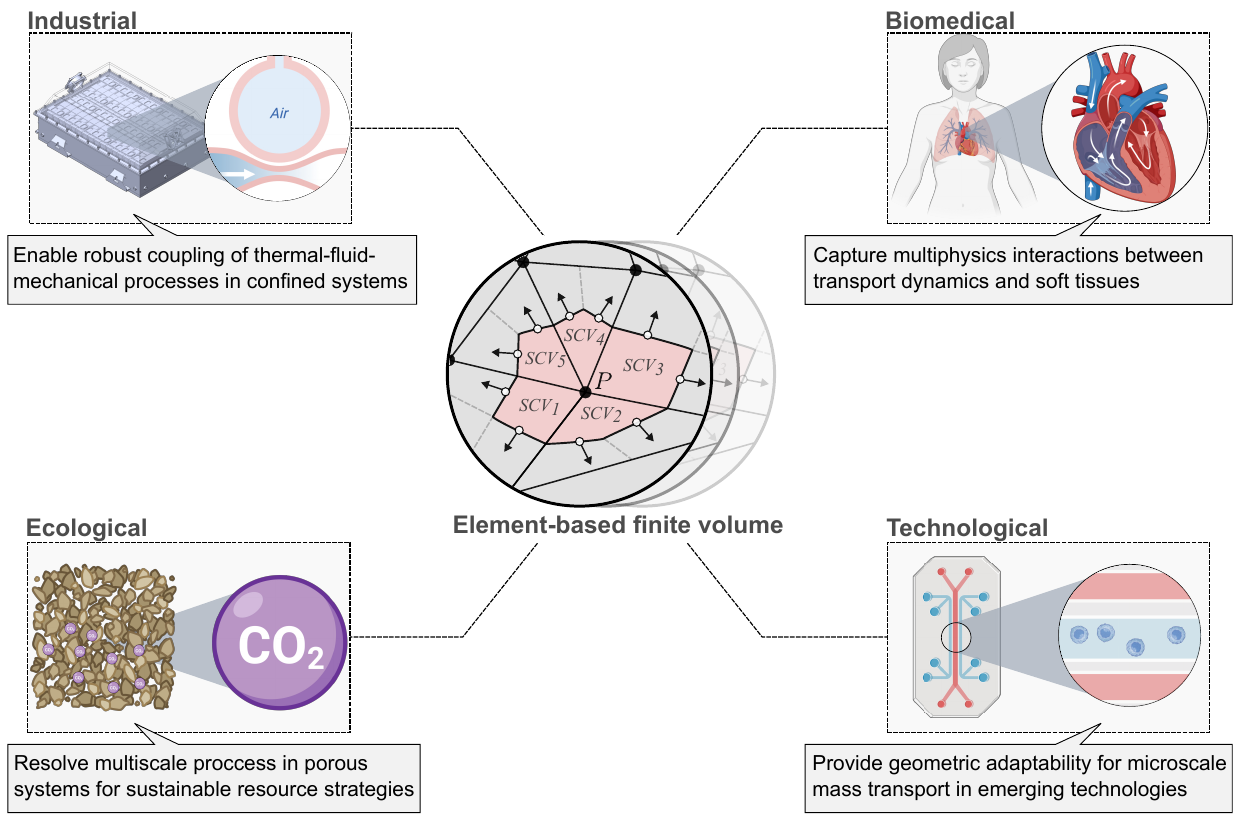}
   \caption{Representative application domains where the Element-based Finite Volume Method offers distinct modelling advantages. The method’s ability to handle unstructured geometries, enforce local conservation and resolve coupled multiphysics makes it ideally suited for diverse industrial, technological, biomedical and ecological systems characterised by complex interfacial dynamics and tightly coupled transport phenomena. Created with \href{https://www.biorender.com}{BioRender}.}   
   \label{fig:EbFVM_Applications}
\end{figure}

To conclude, we have developed and rigorously demonstrated a novel three-dimensional Element-based Finite Volume Method (EbFVM) that unifies geometric adaptability with strict local conservation for the numerical solution of multiphysics transport problems across complex interfacial domains. Through detailed simulations of textured journal bearing systems, the framework has been shown to accurately capture the coupled fluid-solid dynamics arising from intricate surface topographies, validating its robustness and predictive accuracy. The EbFVM departs from conventional finite volume formulations by enabling unstructured, element-conforming discretisations that preserve mass, momentum and energy conservation at the local scale. This capability makes it well suited for problems where geometric irregularity, strong field coupling and multiscale interactions dominate the physical response. Beyond lubrication, the methodology establishes a generalised computational foundation for the simulation of fluid–structure interactions, interfacial transport and coupled thermo-mechanical processes in engineering and natural systems, such as those exemplified in \autoref{fig:EbFVM_Applications}.

Industrial applications present an impactful extension for the EbFVM, particularly in the design and optimisation of high-performance systems such as rotating machinery \cite{dhar2013fluid, Dhanola2020} and electric drivetrains \cite{enang2017modelling, farfan2019tribology}. These systems are often characterised by confined geometries and dynamically evolving interfaces, operating under severe thermo-mechanical conditions. For example, in turbomachinery, the interaction between rotating blades and coolant flows introduces complex fluid–solid coupling \cite{Campbell2011}, pressure transients and heat transfer challenges across highly curved and discontinuous surfaces, all of which are conditions that are ideally suited to the geometric flexibility and local conservativenes of the EbFVM. Similarly, efficient thermal management is critical in battery systems deployed in electric vehicles for ensuring safety, performance and longevity \cite{Chavan2023}. Here, the EbFVM can be deployed to model conjugate heat transfer in intricate cooling architectures, such as liquid-cooled plates, micro-channel heat sinks and phase-change materials embedded within constrained enclosures \cite{Zhang2005}. These configurations necessitate accurate resolution of multiple physiochemical phenomena, such as heat conduction, forced convection and structural compliance, across topologically irregular interfaces. Traditional discretisation methods often struggle to maintain both accuracy and stability in such settings, while conventional simulation tools capable of handling these challenges tend to incur significant computational overhead, \emph{i.e}, open-source CFD models.

From a technological standpoint, the EbFVM framework can drive advancements in microfluidic platforms and smart bio-integrated systems, particularly those governed by coupled fluid-structure dynamics underpinning drug testing and delivery  \cite{nazary2022application, leung2022guide}. In lab-on-a-chip devices \cite{nan2024development}, fluid transport is often governed by intricate interactions including electrokinetic forces, thermocapillary effects, surface tension gradients and flexible boundaries, typically materialising within sub-millimetre-scale geometries \cite{hu2007multiscale, glatzel2008computational}. The EbFVM’s geometric flexibility and conservative flux formulation allows for such tightly coupled phenomena to be resolved on unstructured and evolving meshes. This enables the accurate prediction of flow dynamics in highly miniaturised and deformable domains, where traditional numerical methods either lack stability or require excessive computational resources. In the context of advanced wound care, the EbFVM could play a transformative role in modelling the multiphysics behaviour of bioactive wound dressings and smart bandages \cite{derakhshandeh2018smart, jiang2023wireless}. These materials often involve simultaneous fluid absorption, heat dissipation, mechanical deformation and biochemical transport through porous media \cite{wang2024chronic}. The method’s capability to handle various multiscale domains governed by sharp property gradients and deforming interfaces allows for accurate predictions of fluid uptake, drug delivery kinetics and thermal regulation. As a result, these modelling capabilites can inform the design of next-generation wound care technologies tailored to accelerate healing, minimise infection risk and revamp therapeutic outcomes for patients.

In the biomedical domain, the EbFVM framework can be employed for modelling the multiphysics interactions between deformable tissues and fluid flow. One illustrative example is cartilage rehydration in orthopaedic biomechanics \cite{moore2017tribological}, where fluid transport and poroelastic deformation dynamics are mutually dependent. Capturing these interactions is essential for understanding nutrient delivery, tissue degeneration and the progression of joint disorders such as osteoarthritis. The EbFVM enables precise resolution of these mechanisms, even in the presence of irregular and evolving tissue morphologies. Another example is cardiac circulation \cite{quarteroni2017integrated}, where the ability to simulate the bidirectional coupling between pulsatile blood flow and myocardial deformation is key to evaluating cardiovascular health, designing medical implants and exploring pathological conditions. With its ability to operate on unstructured meshes and accommodate transient, non-linear dynamics, the EbFVM approach provides a robust computational tool for tackling intricate tissue mechanics within the heart and associated vasculature \cite{formaggia2010cardiovascular}.

Furthermore, environmental and bio-inspired applications represent another field where the proposed EbFVM formulations can offer transformative modelling capabilities. In the realm of carbon capture and storage (CCS) \cite{khan2024geomechanical}, modelling the injection and subsurface migration of supercritical $\mathrm{CO_2}$ through porous rock formations requires precise resolution of multiphase flow dynamics, chemical reactions and spatiotemporal evolution of complex pore-scale interfaces \cite{meigel2022dispersive}. The EbFVM’s ability to handle unstructured meshes while rigorously conserving mass-flow allows it to simulate the long-term evolution of subsurface sequestration under dynamic conditions. The deployment of the EbFVM is equally promising in the design and modelling of bio-inspired materials such as synthetic analogues of fish skin and reptile scales \cite{kong2024bioinspired}, which are increasingly being utilised in advanced applications ranging from soft robotics to protective coatings \cite{rus2015design, su2016bioinspired, wang2025multiscale}. These materials often exhibit hierarchical architectures, spatially varying mechanical properties and moisture-responsive behaviour. Therefore, the EbFVM framework provides the required geometric flexibility and multi-region accuracy to capture these interactions, offering valuable insights into how microstructural characteristics influence macroscopic system responses.

\section{Author's Contributions}
\textbf{S. Ardah:} Writing – review \& editing, Writing – original draft, Methodology, Investigation, Data curation, Conceptualization. \textbf{F.J. Profito:} Writing – review \& editing, Supervision, Methodology, Conceptualization. \textbf{D. Dini:} Writing – review \& editing, Supervision, Resources, Project administration, Methodology, Conceptualization.

\section{Acknowledgments}
S. Ardah and D. Dini would like to acknowledge the support received from the Engineering and Physical Sciences Research Council, United Kingdom (EPSRC) via the InFUSE Prosperity Partnership EP/V038044/1. F.J. Profito would like to acknowledge the funding received from the São Paulo Research Foundation (FAPESP) via grant \#2022/03110-0. D. Dini would also like to acknowledge the funding received from the Engineering and Physical Sciences Research Council, United Kingdom (EPSRC) via grants EP/N025954/1 and EP/V038044/1, as well as the support provided by the Shell/RAEng Research Chair in Complex Engineering Interfaces (RCSRF2122-14-143).

\section{Availability of Data and Materials}
The data that support the findings of this study are available from the corresponding author upon reasonable request or at tribology@imperial.ac.uk.

\section{Conflict of Interest}
The authors declare that they have no known competing financial interests or personal relationships that could have appeared to influence the work reported in this paper.


\newpage
\appendix
\newpage

\section{Three-Dimensional Element Formulations for EbFVM Discretisation}
\label{sec:EbFVM_Discretisation}
\begin{figure}
   \centering
   \includegraphics[width=\linewidth]{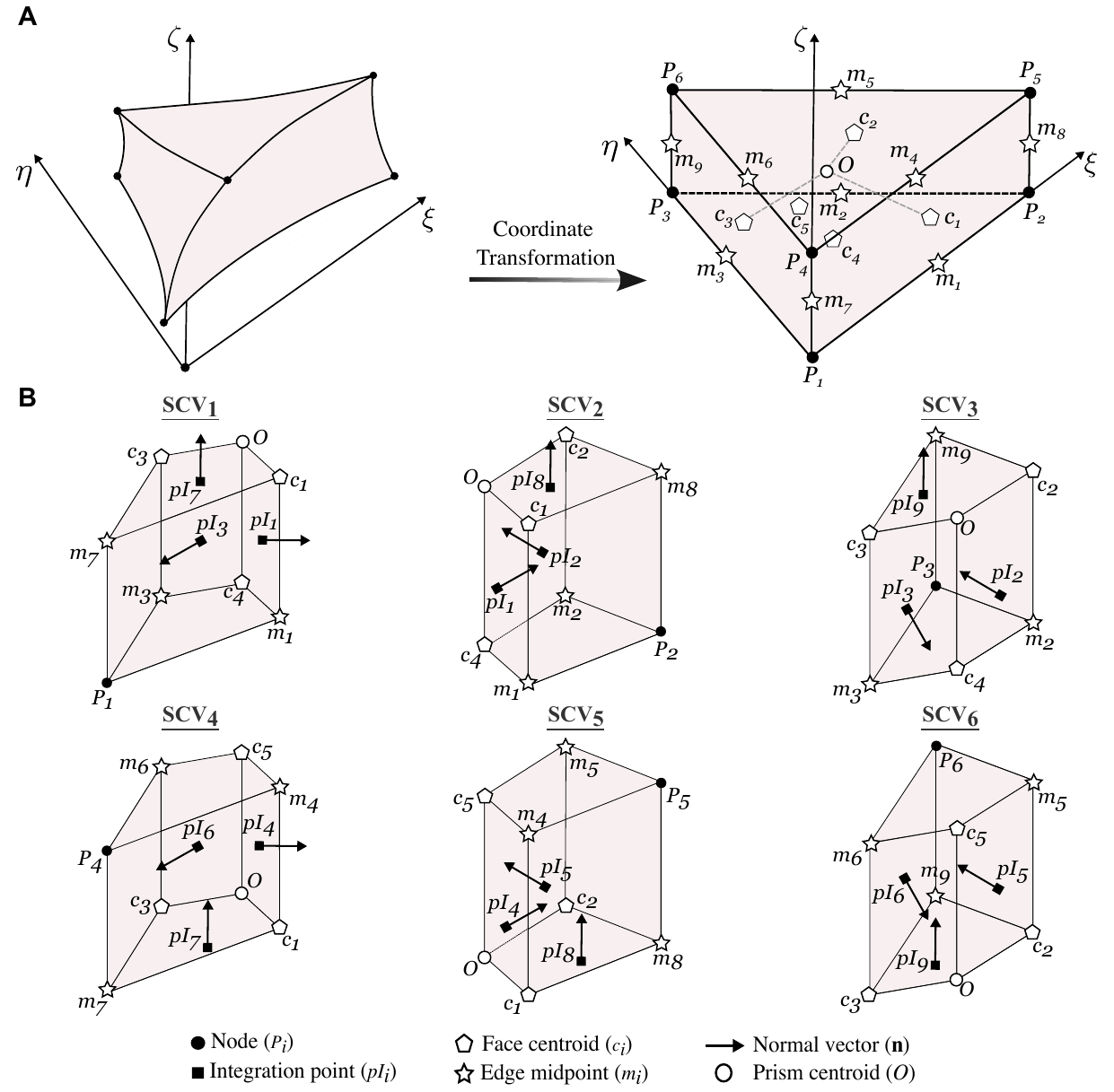}
   \caption{Schematic illustration of the geometric construction and sub-control volume (SCV) generation for the triangular prism element within the EbFVM framework. (\textbf{A}) Transformation of a distorted six-node triangular prism from the physical domain to the standard reference domain, highlighting key geometric features including nodal positions ($P_i$), edge midpoints ($m_i$), face centroids ($c_i$) and element centroid ($O$). (\textbf{B}) Subdivision of the transformed triangular prism into six sub-control volumes (SCVs) using the identified geometric entities. The diagram details the location of integration points ($pI_i$) on SCV faces and the corresponding outward-pointing normal vectors (${\boldsymbol{\mathrm{n}}}_i$). $\mathrm{SCV_1}$ to $\mathrm{SCV_3}$ are positioned at the lower level prismatic element, while $\mathrm{SCV_4}$ to $\mathrm{SCV_6}$ lie at the upper level, collectively ensuring full volumetric coverage around the associated node for local flux evaluation.}
   \label{fig:Triangular_SCVs}
\end{figure}

\begin{figure}
   \centering
   \includegraphics[width=\linewidth]{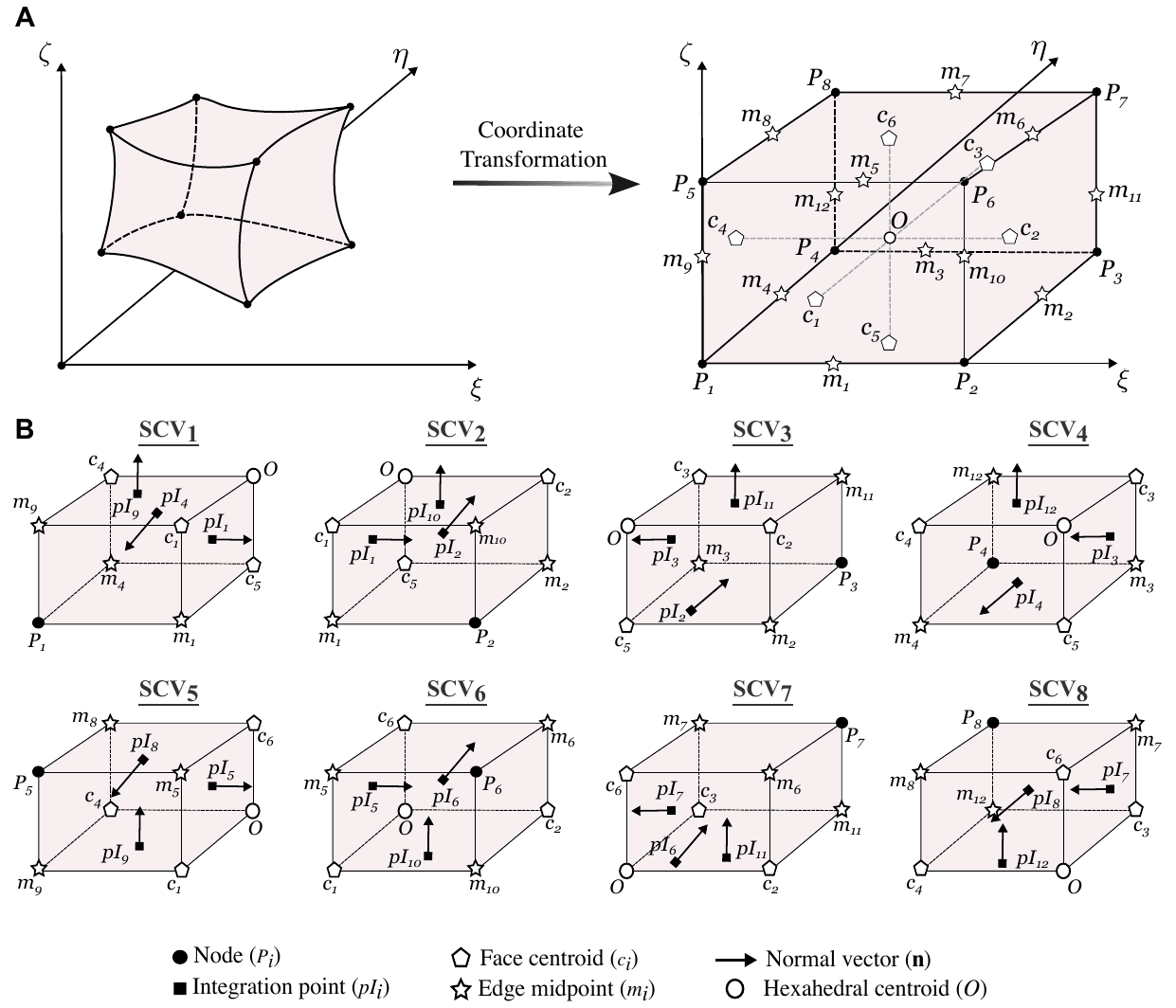}
   \caption{Schematic illustration of the geometric construction and sub-control volume (SCV) generation for the hexahedral element within the EbFVM framework. (\textbf{A}) Transformation of a distorted eight-node hexahedral from the physical domain to the standard reference domain, highlighting key geometric features including nodal positions ($P_i$), edge midpoints ($m_i$), face centroids ($c_i$) and element centroid ($O$). (\textbf{B}) Subdivision of the transformed hexahedral into eight sub-control volumes (SCVs) using the identified geometric entities. The diagram details the location of integration points ($pI_i$) on SCV faces and the corresponding outward-pointing normal vectors (${\boldsymbol{\mathrm{n}}}_i$). $\mathrm{SCV_1}$ to $\mathrm{SCV_4}$ are positioned at the lower level of the hexahedral element, while $\mathrm{SCV_5}$ to $\mathrm{SCV_8}$ lie at the upper level of the hexahedral element, collectively ensuring full volumetric coverage around the associated node for local flux evaluation.}
   \label{fig:Hexahedral_SCVs}
\end{figure}

This section presents the foundational geometrical and mathematical framework underpinning the three-dimensional Element-based Finite Volume Method (EbFVM). The formulation is developed for two primary element types: triangular prisms (\autoref{fig:Triangular_SCVs}) and hexahedra (\autoref{fig:Hexahedral_SCVs}). Each element is defined within a canonical reference (or parametric) domain using a local coordinate system $(\xi, \eta, \zeta)$, which enables systematic mapping to the physical domain through established coordinate transformations. By conducting the discretisation in the reference space, the method ensures geometric robustness, interpolation consistency and numerical stability, even in the presence of mesh irregularities or element distortion. The tensorial expressions and integral transformations described here preserve the conservation properties central to the EbFVM, while supporting accurate approximations of transport phenomena across complex geometries.

While this section focuses on three-dimensional formulations, readers interested in the corresponding two-dimensional derivations for triangular and quadrilateral elements are referred to the original development in \cite{Profito2015Sep}.

\subsection{Trilinear Triangular Prism Element}

\begin{itemize}
    \item Coordinates of the element nodes
        \begin{align}
            \begin{array}{c c c}
                P_1 = (0,0,0) & P_2 = (1,0,0) & P_3 = (0,1,0) \\[3ex]
                P_4 = (0,0,1) & P_5 = (1,0,1) & P_6 = (0,1,1)
            \end{array}
    \end{align}
\end{itemize}

\begin{itemize}
    \item Coordinates of the edges' midpoints
        \begin{align}
            \begin{array}{c c c}
                m_1 = \left(\dfrac{1}{2},0,0\right) & m_2 = \left(\dfrac{1}{2},\dfrac{1}{2},0\right) & m_3 = \left(0,\dfrac{1}{2},0\right) \\[3ex]
                m_4 = \left(\dfrac{1}{2},0,1\right) & m_5 = \left(\dfrac{1}{2},\dfrac{1}{2},1\right) & m_6 = \left(0,\dfrac{1}{2},1\right) \\[3ex]
                m_7 = \left(0,0,\dfrac{1}{2}\right) & m_8 = \left(1,0,\dfrac{1}{2}\right) & m_9 = \left(0,1,\dfrac{1}{2}\right)
            \end{array}
    \end{align}
\end{itemize}

\begin{itemize}
    \item Coordinates of the faces' centroids
        \begin{align}
            \begin{array}{c c c}
                c_1 = \left(\dfrac{1}{2},0,\dfrac{1}{2}\right) & c_2 = \left(\dfrac{1}{2},\dfrac{1}{2},\dfrac{1}{2}\right) & c_3 = \left(0,\dfrac{1}{2},\dfrac{1}{2}\right) \\[3ex]
                c_4 = \left(\dfrac{1}{3},\dfrac{1}{3},0\right) & c_5 = \left(\dfrac{1}{3},\dfrac{1}{3},1\right)
            \end{array}
    \end{align}
\end{itemize}

\begin{itemize}
    \item Coordinates of the integration points
        \begin{align}
            \begin{array}{c c c}
                pI_1 = \left(\dfrac{5}{12},\dfrac{1}{6},\dfrac{1}{4}\right) & 
                pI_2 = \left(\dfrac{5}{12},\dfrac{5}{12},\dfrac{1}{4}\right) & 
                pI_3 = \left(\dfrac{1}{6},\dfrac{5}{12},\dfrac{1}{4}\right) \\[3ex]
                pI_4 = \left(\dfrac{5}{12},\dfrac{1}{6},\dfrac{3}{4}\right) & 
                pI_5 = \left(\dfrac{5}{12},\dfrac{5}{12},\dfrac{3}{4}\right) & 
                pI_6 = \left(\dfrac{1}{6},\dfrac{5}{12},\dfrac{3}{4}\right) \\[3ex]
                pI_7 = \left(\dfrac{5}{24},\dfrac{5}{24},\dfrac{1}{2}\right) & 
                pI_8 = \left(\dfrac{7}{12},\dfrac{5}{24},\dfrac{1}{2}\right) & 
                pI_9 = \left(\dfrac{5}{24},\dfrac{7}{12},\dfrac{1}{2}\right)
            \end{array}
        \end{align}
\end{itemize}

\begin{itemize}
    \item Coordinates of the SCV centroids
        \begin{align}
            \begin{array}{c c c}
                G_1 = \left(\dfrac{5}{24},\dfrac{5}{24},\dfrac{1}{4}\right) & 
                G_2 = \left(\dfrac{7}{12},\dfrac{5}{24},\dfrac{1}{4}\right) & 
                G_3 = \left(\dfrac{5}{24},\dfrac{7}{12},\dfrac{1}{4}\right) \\[3ex]
                G_4 = \left(\dfrac{5}{24},\dfrac{5}{24},\dfrac{3}{4}\right) & 
                G_5 = \left(\dfrac{7}{12},\dfrac{5}{24},\dfrac{3}{4}\right) & 
                G_6 = \left(\dfrac{5}{24},\dfrac{7}{12},\dfrac{3}{4}\right)
            \end{array}
        \end{align}
\end{itemize}

\begin{itemize}
    \item SCV volumes
        \begin{align}
            \begin{array}{c c c}
                V_{SCV_{1}} = \dfrac{1}{12} \mathrm{det} \big[ \boldsymbol{\mathrm{J}} (G_1) \big] & 
                V_{SCV_{2}} = \dfrac{1}{12} \mathrm{det} \big[ \boldsymbol{\mathrm{J}} (G_2) \big] & 
                V_{SCV_{3}} = \dfrac{1}{12} \mathrm{det} \big[ \boldsymbol{\mathrm{J}} (G_3) \big] \\[3ex]
                V_{SCV_{4}} = \dfrac{1}{12} \mathrm{det} \big[ \boldsymbol{\mathrm{J}} (G_4) \big] & 
                V_{SCV_{5}} = \dfrac{1}{12} \mathrm{det} \big[ \boldsymbol{\mathrm{J}} (G_5) \big] & 
                V_{SCV_{6}} = \dfrac{1}{12} \mathrm{det} \big[ \boldsymbol{\mathrm{J}} (G_6) \big]
            \end{array}
        \end{align}
\end{itemize}

\begin{itemize}
    \item Inflow and outflow configurations for the SCVs
        \begin{equation}
        \begin{aligned}
            \text{SCV}_1: &\quad \text{Inflow: } pI_3           \qquad\quad \text{Outflow: } pI_1,\; pI_7 \\
            \text{SCV}_2: &\quad \text{Inflow: } pI_1           \qquad\quad \text{Outflow: } pI_2,\; pI_8 \\
            \text{SCV}_3: &\quad \text{Inflow: } pI_2           \qquad\quad \text{Outflow: } pI_3,\; pI_9 \\
            \text{SCV}_4: &\quad \text{Inflow: } pI_6,\; pI_7   \quad \text{Outflow: } pI_4 \\
            \text{SCV}_5: &\quad \text{Inflow: } pI_4,\; pI_8   \quad \text{Outflow: } pI_5 \\
            \text{SCV}_6: &\quad \text{Inflow: } pI_5,\; pI_9   \quad \text{Outflow: } pI_6
        \end{aligned}
        \label{eq:TriangularPrism_Inflow_Outflow}
        \end{equation}
\end{itemize}

\begin{itemize}
    \item Coordinates of the normal vectors (physical domain) \\[1ex]
    Each internal face of an SCV is formed by four vertices and is treated as a quadrilateral surface. The corresponding face area vector $\Delta \boldsymbol{\mathrm{S}}_f$ and its outward unit normal vector ${\boldsymbol{\mathrm{n}}}^*$ are calculated using a standard surface integration procedure. For a face defined by vertices $\boldsymbol{\mathrm{r}}_1$, $\boldsymbol{\mathrm{r}}_2$, $\boldsymbol{\mathrm{r}}_3$, and $\boldsymbol{\mathrm{r}}_4$, the local area vector is obtained by computing the cross product between two edge vectors originating from a common vertex:
    \begin{equation}
        \Delta {\boldsymbol{\mathrm{{S}}}}_f = \left( {\boldsymbol{\mathrm{{r}}}}_2 - {\boldsymbol{\mathrm{{r}}}}_1 \right) \times \left( {\boldsymbol{\mathrm{{r}}}}_3 - {\boldsymbol{\mathrm{{r}}}}_1 \right),
    \end{equation}
    where the scalar magnitude $\norm{\Delta \boldsymbol{\mathrm{S}}_f}$ represents the surface area of the face and the direction corresponds to the outward normal orientation. The unit normal vector is then calculated as follows:
    \begin{equation}
        {\boldsymbol{\mathrm{n}}}^* = \frac{\Delta \boldsymbol{\mathrm{S}}_f}{\norm{\Delta \boldsymbol{\mathrm{S}}_f}}.
    \end{equation}
\end{itemize}

\begin{itemize}
    \item Interpolation and shape functions
        \begin{align}
            \begin{aligned}
                N_1 (\xi,\eta,\zeta) &= (1-\xi-\eta)(1-\zeta) & \quad
                N_2 (\xi,\eta,\zeta) &= \xi(1-\zeta) & \quad
                N_3 (\xi,\eta,\zeta) &= \eta(1-\zeta) \\[3ex]
                N_4 (\xi,\eta,\zeta) &= \zeta(1-\xi-\eta) & \quad
                N_5 (\xi,\eta,\zeta) &= \xi \zeta & \quad
                N_6 (\xi,\eta,\zeta) &= \eta \zeta
            \end{aligned}
        \end{align}
\end{itemize}

\begin{itemize}
    \item Matrix for gradients interpolation
        \begin{align}
            \boldsymbol{\mathrm{B}}(\xi, \eta, \zeta) = 
            \begin{bmatrix}
                \dfrac{\partial N_1(\xi,\eta,\zeta)}{\partial \xi} & \dfrac{\partial N_1(\xi,\eta,\zeta)}{\partial \eta} & \dfrac{\partial N_1(\xi,\eta,\zeta)}{\partial \zeta} \\[3ex]
                \dfrac{\partial N_2(\xi,\eta,\zeta)}{\partial \xi} & \dfrac{\partial N_2(\xi,\eta,\zeta)}{\partial \eta} & \dfrac{\partial N_2(\xi,\eta,\zeta)}{\partial \zeta} \\[3ex]
                \dfrac{\partial N_3(\xi,\eta,\zeta)}{\partial \xi} & \dfrac{\partial N_3(\xi,\eta,\zeta)}{\partial \eta} & \dfrac{\partial N_3(\xi,\eta,\zeta)}{\partial \zeta} \\[3ex]
                \dfrac{\partial N_4(\xi,\eta,\zeta)}{\partial \xi} & \dfrac{\partial N_4(\xi,\eta,\zeta)}{\partial \eta} & \dfrac{\partial N_4(\xi,\eta,\zeta)}{\partial \zeta} \\[3ex]
                \dfrac{\partial N_5(\xi,\eta,\zeta)}{\partial \xi} & \dfrac{\partial N_5(\xi,\eta,\zeta)}{\partial \eta} & \dfrac{\partial N_5(\xi,\eta,\zeta)}{\partial \zeta} \\[3ex]
                \dfrac{\partial N_6(\xi,\eta,\zeta)}{\partial \xi} & \dfrac{\partial N_6(\xi,\eta,\zeta)}{\partial \eta} & \dfrac{\partial N_6(\xi,\eta,\zeta)}{\partial \zeta}
            \end{bmatrix}
            =
            \begin{bmatrix}
                -(1-\zeta) & -(1-\zeta) & -(1-\xi-\eta) \\[3ex]
                (1-\zeta)  & 0          & -\xi \\[3ex]
                0          & (1-\zeta)  & -\eta \\[3ex]
                -\zeta     & -\zeta     & (1-\xi-\eta) \\[3ex]
                \zeta      & 0          & \xi \\[3ex]
                0          & \zeta      & \eta
            \end{bmatrix}
        \end{align}
\end{itemize}

\begin{itemize}
    \item Jacobian matrix
        \begin{subequations}
            \begin{align}
                \begin{aligned}
                    \boldsymbol{\mathrm{J}} (\xi,\eta,\zeta) = \boldsymbol{\mathrm{Z}}_e \boldsymbol{\mathrm{B}} (\xi,\eta,\zeta)
                \end{aligned}
            \end{align}
            where \( \boldsymbol{\mathrm{Z}}_e \) is the matrix containing the global coordinates of the element nodes:
            \begin{align}
                \begin{aligned}
                    \boldsymbol{\mathrm{Z}}_e = 
                    \begin{bmatrix}
                        x_1 & x_2  & x_3  & x_4  & x_5  & x_6 \\[1ex] 
                        y_1 & y_2  & y_3  & y_4  & y_5  & y_6 \\[1ex]
                        z_1 & z_2  & z_3  & z_4  & z_5  & z_6
                    \end{bmatrix}
                \end{aligned}
            \end{align}
        \end{subequations}
\end{itemize}

\begin{itemize}
    \item Tensor for gradient's interpolation
        \begin{align}
            G_{jk} (\xi,\eta,\zeta) \Rightarrow \boldsymbol{\mathrm{G}} (\xi,\eta,\zeta) = \boldsymbol{\mathrm{B}} (\xi,\eta,\zeta) \; \boldsymbol{\mathrm{J}}^{-1} (\xi,\eta,\zeta)
            \label{eq:G_prism}
        \end{align}
\end{itemize}

\begin{itemize}
    \item {Tensor for diffusion calculation} 
    \begin{subequations}
        \begin{align}
            B_{mjki} &\Rightarrow 
            \begin{cases}
                B_{mjki1} = H_{mjki} (pI_1) - H_{mjki} (pI_3) + H_{mjki} (pI_7) \quad \rightarrow \quad \mathrm{SCV}_1 \\[1ex]
                B_{mjki2} = H_{mjki} (pI_2) - H_{mjki} (pI_1) + H_{mjki} (pI_8) \quad \rightarrow \quad \mathrm{SCV}_2 \\[1ex]
                B_{mjki3} = H_{mjki} (pI_3) - H_{mjki} (pI_2) + H_{mjki} (pI_9) \quad \rightarrow \quad \mathrm{SCV}_3 \\[1ex]
                B_{mjki4} = H_{mjki} (pI_4) - H_{mjki} (pI_6) - H_{mjki} (pI_7) \quad \rightarrow \quad \mathrm{SCV}_4 \\[1ex]
                B_{mjki5} = H_{mjki} (pI_5) - H_{mjki} (pI_4) - H_{mjki} (pI_8) \quad \rightarrow \quad \mathrm{SCV}_5 \\[1ex]
                B_{mjki6} = H_{mjki} (pI_6) - H_{mjki} (pI_5) - H_{mjki} (pI_9) \quad \rightarrow \quad \mathrm{SCV}_6
            \end{cases}
            \label{eq:B_mjki}
        \end{align}
        \noindent where the auxiliary tensor \( H_{mjki} \) is defined as:
        \begin{align}
            H_{mjki} (\xi,\eta,\zeta) = N_m (\xi,\eta,\zeta) \; G_{jk} (\xi,\eta,\zeta) \; \boldsymbol{\mathrm{n}}_i^* (\xi,\eta,\zeta)
            \label{eq:H_prism}
        \end{align}
    \end{subequations}
\end{itemize}

\subsection{Trilinear Hexahedral Element}

\begin{itemize}
    \item Coordinates of the element nodes
        \begin{align}
            \begin{array}{c c c c}
                P_1 = (0,0,0) & P_2 = (1,0,0) & P_3 = (1,1,0) & P_4 = (0,1,0) \\[3ex]
                P_5 = (0,0,1) & P_6 = (1,0,1) & P_7 = (1,1,1) & P_8 = (0,1,1)
            \end{array}
    \end{align}
\end{itemize}

\begin{itemize}
    \item Coordinates of the edges' midpoints
        \begin{align}
            \begin{array}{c c c c}
                m_1 =  \left(\dfrac{1}{2},0,0\right) & m_2 =  \left(1,\dfrac{1}{2},0\right) & m_3 = \left(\dfrac{1}{2},1,0\right) & m_4 =  \left(0,\dfrac{1}{2},0\right)  \\[3ex]
                m_5 =  \left(\dfrac{1}{2},0,1\right) & m_6 =  \left(1,\dfrac{1}{2},1\right) & m_7 =  \left(\dfrac{1}{2},1,1\right) & m_8 =  \left(0,\dfrac{1}{2},1\right)  \\[3ex]
                m_9 =  \left(0,0,\dfrac{1}{2}\right) & m_{10} =  \left(1,0,\dfrac{1}{2}\right) & m_{11} =  \left(1,1,\dfrac{1}{2}\right) & m_{12} =  \left(0,1,\dfrac{1}{2}\right) 
            \end{array}
    \end{align}
\end{itemize}

\begin{itemize}
    \item Coordinates of the faces' centroids
        \begin{align}
            \begin{array}{c c c}
                c_1 =  \left(\dfrac{1}{2},0,\dfrac{1}{2}\right) & c_2 =  \left(1,\dfrac{1}{2},\dfrac{1}{2}\right)  & c_3 = \left(\dfrac{1}{2},1,\dfrac{1}{2}\right)  \\[3ex]
                c_4 =  \left(0,\dfrac{1}{2},\dfrac{1}{2}\right)  &  c_5 =  \left(\dfrac{1}{2},\dfrac{1}{2},0\right) & c_6 =  \left(\dfrac{1}{2},\dfrac{1}{2},1\right)
            \end{array}
    \end{align}
\end{itemize}

\begin{itemize}
    \item Coordinates of the integration points
        \begin{align}
            \begin{array}{c c c c}
                pI_1 = \left(\dfrac{1}{2},\dfrac{1}{4},\dfrac{1}{4}\right) & 
                pI_2 = \left(\dfrac{3}{4},\dfrac{1}{2},\dfrac{1}{4}\right) & 
                pI_3 = \left(\dfrac{1}{2},\dfrac{3}{4},\dfrac{1}{4}\right) &
                pI_4 = \left(\dfrac{1}{4},\dfrac{1}{2},\dfrac{1}{4}\right) \\[3ex] 
                pI_5 = \left(\dfrac{1}{2},\dfrac{1}{4},\dfrac{3}{4}\right) & 
                pI_6 = \left(\dfrac{3}{4},\dfrac{1}{2},\dfrac{3}{4}\right) &
                pI_7 = \left(\dfrac{1}{2},\dfrac{3}{4},\dfrac{3}{4}\right) & 
                pI_8 = \left(\dfrac{1}{4},\dfrac{1}{2},\dfrac{3}{4}\right) \\[3ex] 
                pI_9 = \left(\dfrac{1}{4},\dfrac{1}{4},\dfrac{1}{2}\right) &
                pI_{10} = \left(\dfrac{3}{4},\dfrac{1}{4},\dfrac{1}{2}\right) & 
                pI_{11} = \left(\dfrac{3}{4},\dfrac{3}{4},\dfrac{1}{2}\right) & 
                pI_{12} = \left(\dfrac{1}{4},\dfrac{3}{4},\dfrac{1}{2}\right)
            \end{array}
        \end{align}
\end{itemize}

\begin{itemize}
    \item Coordinates of the SCV centroids
        \begin{align}
            \begin{array}{c c c c}
                G_1 = \left(\dfrac{1}{4},\dfrac{1}{4},\dfrac{1}{4}\right) & 
                G_2 = \left(\dfrac{3}{4},\dfrac{1}{4},\dfrac{1}{4}\right) & 
                G_3 = \left(\dfrac{3}{4},\dfrac{3}{4},\dfrac{1}{4}\right) &
                G_4 = \left(\dfrac{1}{4},\dfrac{3}{4},\dfrac{1}{4}\right) \\[3ex] 
                G_5 = \left(\dfrac{1}{4},\dfrac{1}{4},\dfrac{3}{4}\right) & 
                G_6 = \left(\dfrac{3}{4},\dfrac{1}{4},\dfrac{3}{4}\right) &
                G_7 = \left(\dfrac{3}{4},\dfrac{3}{4},\dfrac{3}{4}\right) &
                G_8 = \left(\dfrac{1}{4},\dfrac{3}{4},\dfrac{3}{4}\right)
            \end{array}
        \end{align}
\end{itemize}

\begin{itemize}
    \item SCV volumes
        \begin{align}
            \begin{array}{c c c c}
                V_{SCV_{1}} = \dfrac{1}{8} \mathrm{det} \big[ \boldsymbol{\mathrm{J}} (G_1) \big] & 
                V_{SCV_{2}} = \dfrac{1}{8} \mathrm{det} \big[ \boldsymbol{\mathrm{J}} (G_2) \big] & 
                V_{SCV_{3}} = \dfrac{1}{8} \mathrm{det} \big[ \boldsymbol{\mathrm{J}} (G_3) \big] &
                V_{SCV_{4}} = \dfrac{1}{8} \mathrm{det} \big[ \boldsymbol{\mathrm{J}} (G_4) \big] \\[3ex] 
                V_{SCV_{5}} = \dfrac{1}{8} \mathrm{det} \big[ \boldsymbol{\mathrm{J}} (G_5) \big] & 
                V_{SCV_{6}} = \dfrac{1}{8} \mathrm{det} \big[ \boldsymbol{\mathrm{J}} (G_6) \big] & 
                V_{SCV_{7}} = \dfrac{1}{8} \mathrm{det} \big[ \boldsymbol{\mathrm{J}} (G_7) \big] & 
                V_{SCV_{8}} = \dfrac{1}{8} \mathrm{det} \big[ \boldsymbol{\mathrm{J}} (G_8) \big] 
            \end{array}
        \end{align}
\end{itemize}

\begin{itemize}
    \item Inflow and outflow configurations for the SCVs
        \begin{equation}
        \begin{aligned}
            \text{SCV}_1: &\quad \text{Inflow: } pI_4 \qquad\quad\quad \text{Outflow: } pI_1,\; pI_9 \\
            \text{SCV}_2: &\quad \text{Inflow: } pI_1 \qquad\quad\quad \text{Outflow: } pI_2,\; pI_{10} \\
            \text{SCV}_3: &\quad \text{Inflow: } pI_2 \qquad\quad\quad \text{Outflow: } pI_3,\; pI_{11} \\
            \text{SCV}_4: &\quad \text{Inflow: } pI_3 \qquad\quad\quad \text{Outflow: } pI_4,\; pI_{12} \\
            \text{SCV}_5: &\quad \text{Inflow: } pI_8,\; pI_9 \quad\quad \text{Outflow: } pI_5 \\
            \text{SCV}_6: &\quad \text{Inflow: } pI_5,\; pI_{10} \quad\quad \text{Outflow: } pI_6 \\
            \text{SCV}_7: &\quad \text{Inflow: } pI_6,\; pI_{11} \quad\quad \text{Outflow: } pI_7 \\
            \text{SCV}_8: &\quad \text{Inflow: } pI_7,\; pI_{12} \quad\quad \text{Outflow: } pI_8
        \end{aligned}
        \label{eq:Hexahedral_Inflow_Outflow}
        \end{equation}
\end{itemize}

\begin{itemize}
    \item Coordinates of the normal vectors (physical domain) \\[1ex]
    Each internal face of an SCV is formed by four vertices and is treated as a quadrilateral surface. The corresponding face area vector $\Delta \boldsymbol{\mathrm{S}}_f$ and its outward unit normal vector ${\boldsymbol{\mathrm{n}}}^*$ are calculated using a standard surface integration procedure. For a face defined by vertices $\boldsymbol{\mathrm{r}}_1$, $\boldsymbol{\mathrm{r}}_2$, $\boldsymbol{\mathrm{r}}_3$, and $\boldsymbol{\mathrm{r}}_4$, the local area vector is obtained by computing the cross product between two edge vectors originating from a common vertex:
    \begin{equation}
        \Delta {\boldsymbol{\mathrm{{S}}}}_f = \left( {\boldsymbol{\mathrm{{r}}}}_2 - {\boldsymbol{\mathrm{{r}}}}_1 \right) \times \left( {\boldsymbol{\mathrm{{r}}}}_3 - {\boldsymbol{\mathrm{{r}}}}_1 \right),
    \end{equation}
    where the scalar magnitude $\norm{\Delta \boldsymbol{\mathrm{S}}_f}$ represents the surface area of the face and the direction corresponds to the outward normal orientation. The unit normal vector is then calculated as follows:
    \begin{equation}
        {\boldsymbol{\mathrm{n}}}^* = \frac{\Delta \boldsymbol{\mathrm{S}}_f}{\norm{\Delta \boldsymbol{\mathrm{S}}_f}}.
    \end{equation}
\end{itemize}

\begin{itemize}
    \item Interpolation and shape functions:
        \begin{align}
            \begin{aligned}
                N_1 (\xi,\eta,\zeta) &= (1-\xi)(1-\eta)(1-\zeta) & \quad
                N_2 (\xi,\eta,\zeta) &= \xi(1-\eta)(1-\zeta) \\[2ex]
                N_3 (\xi,\eta,\zeta) &= \xi\eta(1-\zeta) & \quad
                N_4 (\xi,\eta,\zeta) &= \eta(1-\xi)(1-\zeta) \\[2ex]
                N_5 (\xi,\eta,\zeta) &= \zeta(1-\xi)(1-\eta) & \quad
                N_6 (\xi,\eta,\zeta) &= \xi\zeta(1-\eta) \\[2ex]
                N_7 (\xi,\eta,\zeta) &= \xi\eta\zeta & \quad
                N_8 (\xi,\eta,\zeta) &= \eta\zeta(1-\xi) 
            \end{aligned}
        \end{align}
\end{itemize}

\begin{itemize}
    \item Matrix for gradients interpolation:
        \begin{align}
            \begin{aligned}
                \boldsymbol{\mathrm{B}}(\xi, \eta, \zeta) &= 
                \begin{bmatrix}
                    \dfrac{\partial N_1(\xi,\eta,\zeta)}{\partial \xi} & \dfrac{\partial N_1(\xi,\eta,\zeta)}{\partial \eta} & \dfrac{\partial N_1(\xi,\eta,\zeta)}{\partial \zeta} \\[3ex]
                    \dfrac{\partial N_2(\xi,\eta,\zeta)}{\partial \xi} & \dfrac{\partial N_2(\xi,\eta,\zeta)}{\partial \eta} & \dfrac{\partial N_2(\xi,\eta,\zeta)}{\partial \zeta} \\[3ex]
                    \dfrac{\partial N_3(\xi,\eta,\zeta)}{\partial \xi} & \dfrac{\partial N_3(\xi,\eta,\zeta)}{\partial \eta} & \dfrac{\partial N_3(\xi,\eta,\zeta)}{\partial \zeta} \\[3ex]
                    \dfrac{\partial N_4(\xi,\eta,\zeta)}{\partial \xi} & \dfrac{\partial N_4(\xi,\eta,\zeta)}{\partial \eta} & \dfrac{\partial N_4(\xi,\eta,\zeta)}{\partial \zeta} \\[3ex]
                    \dfrac{\partial N_5(\xi,\eta,\zeta)}{\partial \xi} & \dfrac{\partial N_5(\xi,\eta,\zeta)}{\partial \eta} & \dfrac{\partial N_5(\xi,\eta,\zeta)}{\partial \zeta} \\[3ex]
                    \dfrac{\partial N_6(\xi,\eta,\zeta)}{\partial \xi} & \dfrac{\partial N_6(\xi,\eta,\zeta)}{\partial \eta} & \dfrac{\partial N_6(\xi,\eta,\zeta)}{\partial \zeta} \\[3ex]
                    \dfrac{\partial N_7(\xi,\eta,\zeta)}{\partial \xi} & \dfrac{\partial N_7(\xi,\eta,\zeta)}{\partial \eta} & \dfrac{\partial N_7(\xi,\eta,\zeta)}{\partial \zeta} \\[3ex]
                    \dfrac{\partial N_8(\xi,\eta,\zeta)}{\partial \xi} & \dfrac{\partial N_8(\xi,\eta,\zeta)}{\partial \eta} & \dfrac{\partial N_8(\xi,\eta,\zeta)}{\partial \zeta}
                \end{bmatrix} \\[3ex]
                &= 
                \begin{bmatrix}
                    - (1-\eta)(1-\zeta) & - (1-\xi)(1-\zeta) & - (1-\xi)(1-\eta) \\[2ex]
                    (1-\eta)(1-\zeta) & - \xi(1-\zeta)     & - \xi(1-\eta) \\[2ex]
                    \eta(1-\zeta)     &  \xi(1-\zeta)     & - \xi \eta \\[2ex]
                    - \eta(1-\zeta)     &  (1-\xi)(1-\zeta) & - \eta(1-\xi) \\[2ex]
                    - \zeta(1-\eta)     & - \zeta(1-\xi)    & (1-\xi)(1-\eta) \\[2ex]
                    \zeta(1-\eta)     & - \xi \zeta       &  \xi(1-\eta) \\[2ex]
                    \eta \zeta        &  \xi \zeta        &  \xi \eta \\[2ex]
                    - \eta \zeta        &  \zeta(1-\xi)     &  \eta(1-\xi)
                \end{bmatrix}
            \end{aligned}
        \end{align}
\end{itemize}

\begin{itemize}
    \item Jacobian matrix:
        \begin{subequations}
            \begin{align}
                \begin{aligned}
                    \boldsymbol{\mathrm{J}} (\xi,\eta,\zeta) = \boldsymbol{\mathrm{Z}}_e \boldsymbol{\mathrm{B}} (\xi,\eta,\zeta)
                \end{aligned}
            \end{align}
            where \( \boldsymbol{Z_e} \) is the matrix containing the global coordinates of the element nodes:
            \begin{align}
                \begin{aligned}
                    \boldsymbol{Z_e} = 
                    \begin{bmatrix}
                        x_1 & x_2 & x_3 & x_4 & x_5 & x_6 & x_7 & x_8 \\[1ex] 
                        y_1 & y_2 & y_3 & y_4 & y_5 & y_6 & y_7 & y_8 \\[1ex]
                        z_1 & z_2 & z_3 & z_4 & z_5 & z_6 & z_7 & z_8
                    \end{bmatrix}
                \end{aligned}
            \end{align}
        \end{subequations}
\end{itemize}

\begin{itemize}
    \item Tensor for gradient's interpolation
        \begin{align}
            G_{jk} (\xi,\eta,\zeta) \Rightarrow \boldsymbol{G} (\xi,\eta,\zeta) = \boldsymbol{B} (\xi,\eta,\zeta) \boldsymbol{J}^{-1} (\xi,\eta,\zeta)
            \label{eq:G_hexa}
        \end{align}
\end{itemize}

\begin{itemize}
    \item {Tensor for diffusion calculation} 
    \begin{subequations}
        \begin{align}
            B_{mjki} &\Rightarrow 
            \begin{cases}
                B_{mjki1} = H_{mjki} (pI_1) - H_{mjki} (pI_4) + H_{mjki} (pI_9) \quad & \rightarrow \quad \mathrm{SCV}_1 \\[1ex]
                B_{mjki2} = H_{mjki} (pI_2) - H_{mjki} (pI_1) + H_{mjki} (pI_{10}) \quad & \rightarrow \quad \mathrm{SCV}_2 \\[1ex]
                B_{mjki3} = H_{mjki} (pI_3) - H_{mjki} (pI_2) + H_{mjki} (pI_{11}) \quad & \rightarrow \quad \mathrm{SCV}_3 \\[1ex]
                B_{mjki4} = H_{mjki} (pI_4) - H_{mjki} (pI_3) + H_{mjki} (pI_{12}) \quad & \rightarrow \quad \mathrm{SCV}_4 \\[1ex]
                B_{mjki5} = H_{mjki} (pI_5) - H_{mjki} (pI_8) - H_{mjki} (pI_9) \quad & \rightarrow \quad \mathrm{SCV}_5 \\[1ex]
                B_{mjki6} = H_{mjki} (pI_6) - H_{mjki} (pI_5) - H_{mjki} (pI_{10}) \quad & \rightarrow \quad \mathrm{SCV}_6 \\[1ex]
                B_{mjki7} = H_{mjki} (pI_7) - H_{mjki} (pI_6) - H_{mjki} (pI_{11}) \quad & \rightarrow \quad \mathrm{SCV}_7 \\[1ex]
                B_{mjki8} = H_{mjki} (pI_8) - H_{mjki} (pI_7) - H_{mjki} (pI_{12}) \quad & \rightarrow \quad \mathrm{SCV}_8
            \end{cases}
            \label{eq:B_hexa}
        \end{align}
        
        \noindent where the auxiliary tensor \( H_{mjki} \) is defined as:
        \begin{align}
            H_{mjki} (\xi,\eta,\zeta) = N_m (\xi,\eta,\zeta) \; G_{jk} (\xi,\eta,\zeta) \; \boldsymbol{\mathrm{n}}_i^* (\xi,\eta,\zeta)
            \label{eq:H_hexa}
        \end{align}
    \end{subequations}
\end{itemize}

\section{Discretisation of the Diffusive Term}
\label{sec:Appendix_Diffusion}
As presented in \autoref{eq:Eq_10}, the diffusive flux at an integration point \( pI \) within a given element \( e \) is expressed as:
\begin{equation}
    {\mathfrak{D}}_e^n(pI) = \left[ \boldsymbol{\Gamma}^d(pI) \nabla \phi(pI) \right]^n_e \cdot \boldsymbol{\mathrm{n}}^*_e(pI),
    \label{eq:DiffFlux_Pointwise}
\end{equation}
where \( pI = (\xi_{pI}, \eta_{pI}, \zeta_{pI}) \) denotes the coordinates of the integration point in the reference domain and ${\boldsymbol{\mathrm{n}}}^*_e(pI)$ is the outward unit normal vector at $pI$. For brevity, the superscripts \( n \) and \( e \), indicating the time level and element index, respectively, are omitted in the derivations that follow.

\subsection{Interpolation of the Diffusivity Tensor}
The diffusivity tensor \( \boldsymbol{\Gamma}^d(pI) \in \mathbb{R}^{3 \times 3} \) at the integration point $pI$ is approximated and expressed component-wise as:
\begin{equation}
    \Gamma^d_{ij}(pI) \approx \sum_{m=1}^M N_m(pI) \, \Gamma^d_{ijm}, \quad i,j = 1,2,3,
    \label{eq:DiffusivityInterp}
\end{equation}
where \( \Gamma^d_{ijm} \) is the nodal value of the \( (i,j) \)-component of the diffusivity tensor at node \( m \), \( M \) is the number of nodes per element, and \( N_m(pI) \) denotes the shape function associated with node \( m \), evaluated at \( pI \). For details on the mapping and properties of \( N_m \), see \ref{sec:EbFVM_Discretisation}.

\subsection{Interpolation of the Gradient of the Transported Field}
The gradient of the transported scalar field \( \phi \) at the integration point \( pI \) is similarly approximated as:
\begin{equation}
    \frac{\partial \phi}{\partial x_j}(pI) \approx \sum_{k=1}^M G_{jk}(pI) \, \phi_k, \quad j = 1,2,3,
    \label{eq:GradientInterp}
\end{equation}
where \( \phi_k \) is the nodal value of the transported field at node \( k \) and \( G_{jk} \) is the gradient interpolation tensor that encodes both the geometric mapping and the derivative of the shape functions in physical space (see \autoref{eq:G_prism} and \autoref{eq:G_hexa}).

\subsection{Approximation of the Diffusive Flux}
Substituting \autoref{eq:DiffusivityInterp} and \autoref{eq:GradientInterp} into \autoref{eq:DiffFlux_Pointwise} yields the approximated  diffusive flux at $pI$ given as:
\begin{equation}
    \mathfrak{D}(pI) \approx \sum_{m,k=1}^M H_{mjki}(pI) \, \Gamma^d_{ijm} \, \phi_k, \quad i,j = 1,2,3,
    \label{eq:PointwiseFluxExpanded}
\end{equation}
where the auxiliary tensor \( H_{mjki} \) is defined by:
\begin{equation}
    H_{mjki}(pI) = N_m(pI) \, G_{jk}(pI) \, n_i^*(pI).
    \label{eq:H_Tensor}
\end{equation}
Note that \( H_{mjki} \) depends only on the element shape functions and geometry, and can therefore be precomputed for fixed meshes to improve computational efficiency (see \autoref{eq:H_prism} and \autoref{eq:H_hexa}).

\subsection{Integration over Sub-Control Volumes}
The total diffusive flux through sub-control volume \( s \) is computed by adding the fluxes over its corresponding integration faces:
\begin{equation}
    \mathfrak{D}_s = \sum_{m,k=1}^M B_{mjkis} \, \Gamma^d_{ijm} \, \phi_k = \mathcal{D}_{ks} \, \phi_k,  \quad i,j = 1,2,3,
    \label{eq:SCV_Diffusion}
\end{equation}
where \( B_{mjkis} \) is the precomputed geometric diffusion tensor for SCV \( s \) (see \autoref{eq:B_mjki} and \autoref{eq:B_hexa}), and \( \mathcal{D}_{ks} \) denotes the local SCV-level diffusion matrix.

\subsection{Assembly of the Element Diffusion Vector}
The vector of total diffusive flux through the integration faces of each SCV within element \(e\) is written compactly as:
\begin{equation}
    \boldsymbol{\mathfrak{D}}_e^n = \boldsymbol{\mathcal{D}}_e^n \, \boldsymbol{\phi}_e^n,
    \label{eq:ElemDiffusionFinal}
\end{equation}
where \( \boldsymbol{\mathfrak{D}}_e^n \in \mathbb{R}^M \) is the element vector of diffusive fluxes through each SCV at time step \( n \), \( \boldsymbol{\phi}_e^n \in \mathbb{R}^M \) is the nodal values of the transported field, and \( \boldsymbol{\mathcal{D}}_e^n \in \mathbb{R}^{M \times M} \) is the element diffusion matrix accounting for the influence of the element’s geometry, material diffusivity and spatial discretisation.
        
\section{Discretisation of the Convective Term}
\label{sec:Appendix_Convection}
The discretisation of the convective term follows a procedure analogous to that used for the diffusive term.  For each integration point $pI = (\xi_{pI}, \eta_{pI}, \zeta_{pI})$ within a given element $e$, the convective flux at time step $n$ is expressed as:
\begin{equation}
        {\mathfrak{C}}_e^{n}(pI) = \left[ \tilde{{q}}(pI) \, \phi(pI) \right]_e^n,
        \qquad \tilde{{q}}_e^n(pI) = \left[ \boldsymbol{\Gamma}^c(pI) \, \boldsymbol{v}(pI) \right]_e^n \cdot \boldsymbol{\mathrm{n}}^*_e(pI),
        \label{eq:ConvectiveFluxPoint}
\end{equation}
where $\tilde{{q}}(pI)$ denotes the local convective mass flux, $\boldsymbol{\Gamma}^c$ is the convective transport tensor, $\boldsymbol{v}$ is the transport velocity field, and $\boldsymbol{\mathrm{n}}^*_e$ is the outward unit normal vector associated with the SCV face at the integration point.

\subsection{Interpolation of the Local Convective Flux}
The local convective flux \(\tilde{{q}}(pI)\) is expressed in tensor index notation as:
\begin{equation}
    \tilde{{q}}(pI) = {\mathcal{C}_i} (pI) {n}^*_i(pI), \quad \quad \mathcal{C}_i(pI) = {\Gamma}^c_{ij}(pI) v_j(pI), \quad \quad i, j = 1, 2, 3.
    \label{Eq:Eq_52}
\end{equation}
Assuming that the components of ${\mathcal{C}_i} (pI)$ are interpolated using the same family of shape functions $N_m(pI)$ adopted for the primary variables, their approximated form becomes:
\begin{equation}
    {\mathcal{C}_i} (pI) \approx N_m(pI) {\mathcal{C}_{im}}, \quad \quad m = 1, 2, ..., M,
\end{equation}
where ${\mathcal{C}_{im}}$ are the nodal values of $\mathcal{C}_i$ and $M$ is the number of nodes per element. Substituting the interpolation functions $N_m$ into \autoref{Eq:Eq_52} yields:
\begin{equation}
    \tilde{{q}}(pI) \approx {Q}_{mi}(pI) \mathcal{C}_{im}, \quad \quad \mathcal{C}_{im} = \left( {\Gamma}^{c}_{ij} v_j \right)_m , \quad \quad {Q}_{mi}(pI) = N_m (pI) n^*_i (pI),
    \label{eq:Eq_57}
\end{equation}
where \( Q_{mi}(pI) \) captures geometric and topological contributions and can be precomputed for fixed meshes.

\subsection{Upwind Scheme and Assembly of the Element Convective Vector}
To approximate the convected scalar field ${\phi}$ at integration points within each element, a directional upwind redistribution scheme is employed. Unlike classical approaches such as the Flow-Weighted Upwind Scheme (FWUS) \cite{Profito2015Sep}, the present formulation constructs a redistribution matrix that is directly informed by the direction and magnitude of the convective flux at each integration point. This ensures local conservation, directional consistency and numerical stability, while preserving simplicity in implementation.

Central to this approach is the formulation of a convective redistribution matrix that systematically differentiates between donor and receiver sub-control volumes based on the direction of the local convective flux. The flux directionality is governed by $\tilde{{q}}(pI)$, evaluated at each integration point $pI = (\xi_{pI}, \eta_{pI}, \zeta_{pI})$, which dictates the orientation of convective flux across the SCV integration faces. Specifically:
\begin{itemize}
        \item If $\tilde{{q}}(pI) > 0$, the convective flux is considered outgoing from the donor SCV, with the integration point functioning as an outflow interface toward the receiver SCV.
        \item If $\tilde{{q}}(pI) < 0$, the flux reverses direction, such that the integration point becomes an inflow interface, and the original receiver SCV now serves as the donor.
\end{itemize}
Each integration point is thus uniquely assigned to a donor–receiver SCV pair, depending on the local flow direction. The inflow–outflow relationships used in this identification are summarised in \autoref{eq:TriangularPrism_Inflow_Outflow} and \autoref{eq:Hexahedral_Inflow_Outflow} for triangular prism and hexahedral elements, respectively.

Given these relationships, the vector of total convective flux through the integration faces of each SCV within element \(e\) is constructed and written compactly as:
\begin{equation}
    \boldsymbol{\mathfrak{C}}_e^n = \boldsymbol{\mathcal{C}}_e^n \, \boldsymbol{\phi}_e^n,
    \label{eq:ElemConvectiveFinal}
\end{equation}
where \( \boldsymbol{\mathfrak{C}}_e^n \in \mathbb{R}^M \) is the element vector of convective fluxes through each SCV at time step \( n \), \( \boldsymbol{\phi}_e^n \in \mathbb{R}^M \) is the nodal values of the transported field, and \( \boldsymbol{\mathcal{C}}_e^n \in \mathbb{R}^{M \times M} \) is the element convective matrix accounting for the influence of the element’s geometry, fluxes directionality and spatial discretisation.

\section{Discretisation of the Source and Discrete Time-Dependent Terms}
\label{sec:Appendix_SourceTerm}
As defined in \autoref{eq:Eq_10}, the discretised scalar transport equation includes three distinct source and time-dependent terms per SCV $s$ within element $e$. These can be written component-wise as:
\begin{gather}
    \begin{aligned} 
            \mathfrak{S}_{1,e,s}^{n} &= 
            \Bigg[
            \frac{(\rho \phi)^n}{\Delta t}
            \Bigg]_{P} \Delta \Omega^{e}_{s}, &\qquad s = 1,2,\dots, M \\
            \mathfrak{S}_{2,e,s}^{n} &=
            -\Bigg[
            \frac{(\rho {\phi})^{n-1}}{\Delta t}
            \Bigg]_P \Delta \Omega^{e}_{s}, &\qquad s = 1,2,\dots, M \\
            \mathfrak{S}_{3,e,s}^{n} &= 
            \bigg[ 
            {Q^{\phi}} \bigg]^{n}_{P} \Delta \Omega^{e}_{s}, &\qquad s = 1,2,\dots, M 
            \label{eq:source_scalar}
    \end{aligned}
\end{gather}
where $\Delta \Omega^{e}_{s}$ denotes the volume of the SCV $s$ within element $e$ that contributes to the CV associated with a node $P$. It is important to note that all terms in \autoref{eq:source_scalar} are evaluated explicitly at the node $P$ connected to the SCV $s$.

To facilitate efficient element-wise assembly, the nodal source contributions are reformulated in vector notation. Each total source contribution $\boldsymbol{\mathfrak{S}}_{i,e}^{n}$, where $i = 1,2,3$, corresponds to a different term in the discretised governing equation and is constructed from the auxiliary vectors $\boldsymbol{\mathcal{S}}_{i,e}^{n}$ and the nodal values of the transported scalar field. These expressions are detailed as follows:
\begin{subequations}
    \begin{gather}
        \begin{aligned} 
                \boldsymbol{{\mathfrak{S}}}_{1,e}^{n} = \boldsymbol{ {\mathcal{S}}}_{1,e}^{n} \cdot \boldsymbol{{{\phi}}}_e^{n} = \underbrace{{\Bigg( \frac{\rho}{\Delta t} \Bigg)^{n}_{P} \Delta \Omega^{e}_{s}}}_{\displaystyle \mathcal{S}^{n}_{1,e,s}} \; {{{\phi}}}_{e,s}^{n},
        \end{aligned}
    \end{gather}
    \begin{gather}
        \begin{aligned} 
                \boldsymbol{{\mathfrak{S}}}_{2,e}^{n} = \boldsymbol{ {\mathcal{S}}}_{2,e}^{n} \cdot \boldsymbol{{{\phi}}}_e^{n-1} = \underbrace{- \Bigg( \frac{\rho}{\Delta t} \Bigg)^{n-1}_{P} \Delta \Omega^{e}_{s}}_{\displaystyle \mathcal{S}^{n}_{2,e,s}} \; \boldsymbol{{{\phi}}}_{e,s}^{n-1},
        \end{aligned}
    \end{gather}
    \begin{gather}
        \begin{aligned} 
                \boldsymbol{{\mathfrak{S}}}_{3,e}^{n} = \boldsymbol{ {\mathcal{S}}}_{3,e}^{n} = \underbrace{{Q^{\phi}}^{n}_{P} \; \Delta \Omega^{e}_{s}}_{\displaystyle \mathcal{S}^{n}_{3,e,s}}.
        \end{aligned}
    \end{gather}
\end{subequations}
In the above expressions, $\boldsymbol{\mathfrak{S}}_{1,e}^{n}$, $\boldsymbol{\mathfrak{S}}_{2,e}^{n}$, and $\boldsymbol{\mathfrak{S}}_{3,e}^{n}$ represent the element vector of source contributions across the sub-control volumes within element $e$, associated respectively with temporal accumulation, time-lagged contributions and external source effects. The auxiliary vectors $\boldsymbol{\mathcal{S}}_{i,e}^{n}$ encode the geometric and physical parameters, such as density $\rho$, control volume size $\Delta \Omega^{e}_{s}$ and source strength $Q^\phi$, while $\boldsymbol{\phi}_e^n$ and $\boldsymbol{\phi}_e^{n-1}$ denote the nodal values of the transported field at time levels $n$ and $n-1$. The operator “$\cdot$” denotes the inner (dot) product applied over the nodal degrees of freedom.

\newpage
\section{Dynamics Solver for Rigid Journal Bearing Systems}
\label{sec:Appendix_Misalignment}
To derive the dynamic equations governing the rigid body motion of the journal relative to the bearing surface, the following kinematic assumptions are adopted:
\begin{enumerate}
    \item \textbf{Small relative displacements and misalignments}: The analysis assumes small-magnitude rigid body displacements of the journal relative to the bearing. Global or gross motion of the system is neglected and, if needed, can be considered separately as inertial forces.    
    \item \textbf{Prescribed Operating Conditions}: Transient effects associated with changes in rotational or sliding speeds are disregarded. Instead, operating speeds are treated as known inputs to the model.
\end{enumerate}

Under these assumptions, the governing equations of motion expressed in the Cartesian coordinate system $xyz$ attached to the bearing centre $B$ (see \autoref{fig:JB_CoordinateSystem}\textbf{A}) are formulated using Newton’s second law as follows:
\begin{align}
        \boldsymbol{\mathrm{M}}_\text{c} \ddot{\boldsymbol{{{\mathrm{q}}}}} (t) + \boldsymbol{{\mathrm{F}}}_\text{c} \left( \boldsymbol{\mathrm{q}}, \Dot{\boldsymbol{{\mathrm{q}}}}, t \right) = \boldsymbol{\mathrm{F}}_\text{ext} (t),
        \label{eq:Rigid_Body_JB}
\end{align}
where $\boldsymbol{\mathrm{M}}_\text{c}$ is the inertia matrix, $\boldsymbol{{\mathrm{F}}}_\text{c}$ is the load vector associated with the hydrodynamic forces and moments arising from the fluid pressure distribution generated within the lubricant film on the journal-bearing interface, and $\boldsymbol{\mathrm{F}}_\text{ext}$ denotes externally applied forces and moments. Moreover, $\boldsymbol{\mathrm{q}}$, $\Dot{\boldsymbol{{\mathrm{q}}}}$ and $\boldsymbol{\ddot{{\mathrm{q}}}}$ are the displacement, velocity and acceleration vectors of the journal relative to the bearing, respectively. The particular definition of these quantities for journal bearing systems is as follows:
\begin{gather}
    \boldsymbol{\mathrm{M}}_\text{c} = 
    \begin{bmatrix}
        m_\text{J} & 0 & 0 & 0
        \\[1ex]
        0 & m_\text{J} & 0 & 0
        \\[1ex]
        0 & 0 & I_\text{X} & 0
        \\[1ex]
        0 & 0 & 0 & I_\text{Y}
    \end{bmatrix},
    \quad \quad
    \boldsymbol{\mathrm{q}} = 
    \begin{bmatrix}
        X_\text{r}
        \\[1ex]
        Y_\text{r}
        \\[1ex]
        A_\text{r}
        \\[1ex]
        B_\text{r}
    \end{bmatrix},
    \quad \quad
    \Dot{\boldsymbol{{\mathrm{q}}}} = 
    \begin{bmatrix}
        \Dot{X}_\text{r}
        \\[1ex]
        \Dot{Y}_\text{r}
        \\[1ex]
        \Dot{A}_\text{r}
        \\[1ex]
        \Dot{B}_\text{r}
    \end{bmatrix},
    \quad \quad
    \ddot{\boldsymbol{{{\mathrm{q}}}}} = 
    \begin{bmatrix}
        \ddot{X}_\text{r}
        \\[1ex]
        \ddot{Y}_\text{r}
        \\[1ex]
        \ddot{A}_\text{r}
        \\[1ex]
        \ddot{B}_\text{r}
    \end{bmatrix} \nonumber
    \\ 
    \boldsymbol{{\mathrm{F}}}_\text{c} = 
    \begin{bmatrix}
        W^{\text{X}}_\text{H}
        \\[1ex]
        W^{\text{Y}}_\text{H}
        \\[1ex]
        M^{\text{X}}_\text{H}
        \\[1ex]
        M^{\text{Y}}_\text{H}
    \end{bmatrix},
    \quad \quad
    \boldsymbol{{\mathrm{F}}}_\text{ext} = 
    \begin{bmatrix}
        F^{\text{X}}_\text{ext}
        \\[1ex]
        F^{\text{Y}}_\text{ext}
        \\[1ex]
        M^{\text{X}}_\text{ext}
        \\[1ex]
        M^{\text{Y}}_\text{ext}
    \end{bmatrix},
\end{gather}
where $m_\text{J}$ is the equivalent journal mass, and $I_\text{X}$ and $I_\text{Y}$ are the moments of inertia of the journal about the BX and BY axes, respectively. Additionally, $X_\text{r}$ and $Y_\text{r}$ refer to the relative displacements of the journal in the BX and BY directions, and $A_\text{r}$ and $B_\text{r}$ denote the relative misalignments around the BX and BY axes. Furthermore, $W$ and $M$ indicate the corresponding forces and moments, respectively.

Considering an implicit numerical integration of \autoref{eq:Rigid_Body_JB}, the following discrete dynamic equation is defined at the time step $n$:
\begin{align}
        \boldsymbol{\mathrm{M}}_\text{c} \ddot{\boldsymbol{{{\mathrm{q}}}}}^{n} + \boldsymbol{{\mathrm{F}}}_\text{c} \left( \boldsymbol{\mathrm{q}}^{n}, \Dot{\boldsymbol{{\mathrm{q}}}}^{n}, t^n \right) = \boldsymbol{{\mathrm{F}}}_\text{ext} (t^n).
        \label{eq:Rigid_Body_JB_2}
\end{align}
This equation can be reformulated as a nonlinear system of equations as follows:
\begin{align}
        \boldsymbol{\Omega} \left( \boldsymbol{\mathrm{q}}^{n} \right) = \boldsymbol{\mathrm{M}_\text{c}} \ddot{\boldsymbol{{{\mathrm{q}}}}}^{n} + {\boldsymbol{{\mathrm{F}}}}_\text{c} \left( \boldsymbol{\mathrm{q}}^{n}, \Dot{\boldsymbol{{\mathrm{q}}}}^{n}, t^n \right) - \boldsymbol{{\mathrm{F}}}_\text{ext} (t^n) = \boldsymbol{\mathrm{0}}.
\end{align}
The operator $\boldsymbol{\Omega}$ denotes the set of nonlinear equations that must be solved at each time step to determine the “\emph{instantaneous equilibrium}” of the journal bearing system. This nonlinear system is resolved using the Newton–Raphson method enhanced by Armijo’s line search strategy, which dynamically adjusts the step length to ensure a sufficient decrease in the residual norm at every iteration \cite{Kelley2003Jan}. The iterative Newton update, derived from the linearised approximation of $\boldsymbol{\Omega} \left(  \boldsymbol{\mathrm{q}}^{\,n} \right)$, is formulated as follows:
\begin{align}
        \boldsymbol{\mathrm{q}}^{n,w} = \boldsymbol{\mathrm{q}}^{n,w-1} + \lambda^{'w} \boldsymbol{\mathrm{s}}^{\,w},
        \quad \quad \quad \quad \quad
        \boldsymbol{\mathcal{D}'} \left( \boldsymbol{\mathrm{q}}^{n,w-1} \right) \boldsymbol{\mathrm{s}}^{w} = -\boldsymbol{\Omega} \left( \boldsymbol{\mathrm{q}}^{n,w-1} \right),
\end{align}
where $\boldsymbol{\mathcal{D}'}$ is the Jacobian matrix (approximated numerically via finite differences), $\boldsymbol{\mathrm{s}}^{w}$ is the Newton increment and $\lambda^{'w}$ is the adaptive step length determined via Armijo’s rule to enforce a monotonic reduction in $\norm{\boldsymbol{\Omega}}$ \cite{Nocedal2006Jun, Dennis0898}. The iterative procedure proceeds until convergence is achieved according to the following relative tolerance criterion:
\begin{equation}
        \left\lVert \boldsymbol{\Omega} \left( \boldsymbol{\mathrm{q}}^{n,w} \right) \right\rVert \leq \varepsilon_\text{NR} \left\lVert \boldsymbol{{\mathrm{F}}}_\text{ext} (t^n) \right\rVert
\end{equation}
where $\varepsilon_\text{NR}$ is a prescribed convergence threshold relative to the magnitude of the external loading. This formulation ensures both robustness and accuracy in solving for the coupled rigid body response of the bearing system.

\newpage
\section{Extended Successive Over Relaxation (SOR) Algorithm for Solving the (\texorpdfstring{$p{-}\theta$}{p-theta}) Generalised Reynolds Equation}
\label{sec:Appendix_p-theta}
\RestyleAlgo{ruled}
\SetKwComment{Comment}{/* }{ */}
\begin{algorithm}[hbt!]
\caption{Pseudo-code for the extended SOR algorithm responsible for solving the modified \( p{-}\theta \) generalised Reynolds equation}
\KwData{ \( \boldsymbol{p}^{n-1}, \boldsymbol{\theta}^{n-1}, \boldsymbol{a}^n, \boldsymbol{b}^n, \boldsymbol{B}^n, p_{\text{cav}}, \omega_p, \omega_\theta, \varepsilon_{SOR}, r_{\max} \)}
\KwResult{Converged \( \boldsymbol{p}^n \) and \( \boldsymbol{\theta}^n \)}
\textbf{Initialisation} \\
\( \bar{p}^{n,0} \gets \bar{p}^{n-1} \), \( \bar{\theta}^{n,0} \gets \bar{\theta}^{n-1} \) \\
\( r \gets 0 \), \( e_{SOR} \gets 1 \)\;
\While{\( e_{SOR} > \varepsilon_{SOR} \) \textbf{and} \( r \leq r_{\max} \)}{
  \( r \gets r + 1 \) \\
  \For{\( P = 1 \) to \( N_T \)}{
    \textbf{Calculate pressure update} \\
    \If{\( {p}^{n,r}_P > p_{\text{cav}} \) \textbf{or} \( {\theta}^{n,r}_P \geq 1 \)}{
      Compute \( P^{n,r}_P \) \\
      \( {p}^{n,r}_P \gets \omega_p {P}^{n,r}_P + (1 - \omega_p) {p}^{n,r}_P \) \\
      \eIf{\( {p}^{n,r}_P \geq p_{\text{cav}} \)}{
        \( {\theta}^{n,r}_P \gets 1 \) \\
      }{
        \( {p}^{n,r}_P \gets p_{\text{cav}} \) \\
      }
    }
    \textbf{Calculate film fraction update} \\
    \If{\( {p}^{n,r}_P \leq p_{\text{cav}} \) \textbf{or} \( {\theta}^{n,r}_P < 1 \)}{
      Compute \( {\theta}^{n,r}_P \) \\
      \( {\theta}^{n,r}_P \gets \omega_\theta {\theta}^{n,r}_P + (1 - \omega_\theta) {\theta}^{n,r}_P \) \\
      \eIf{\( {\theta}^{n,r}_P < 1 \)}{
        \( {p}^{n,r}_P \gets p_{\text{cav}} \) \\
      }{
        \( {\theta}^{n,r}_P \gets 1 \) \\
      }
    }
  }
  \textbf{Compute relative error} \( e_{SOR} \) \\
}
\end{algorithm}

\newpage
\section{Thermo-Hydrodynamic Performance of the Non-Textured Journal Bearing}
\label{sec:Appendix_JBR_NoTexture}
\begin{figure}[H]
   \centering
   \includegraphics[width=\linewidth]{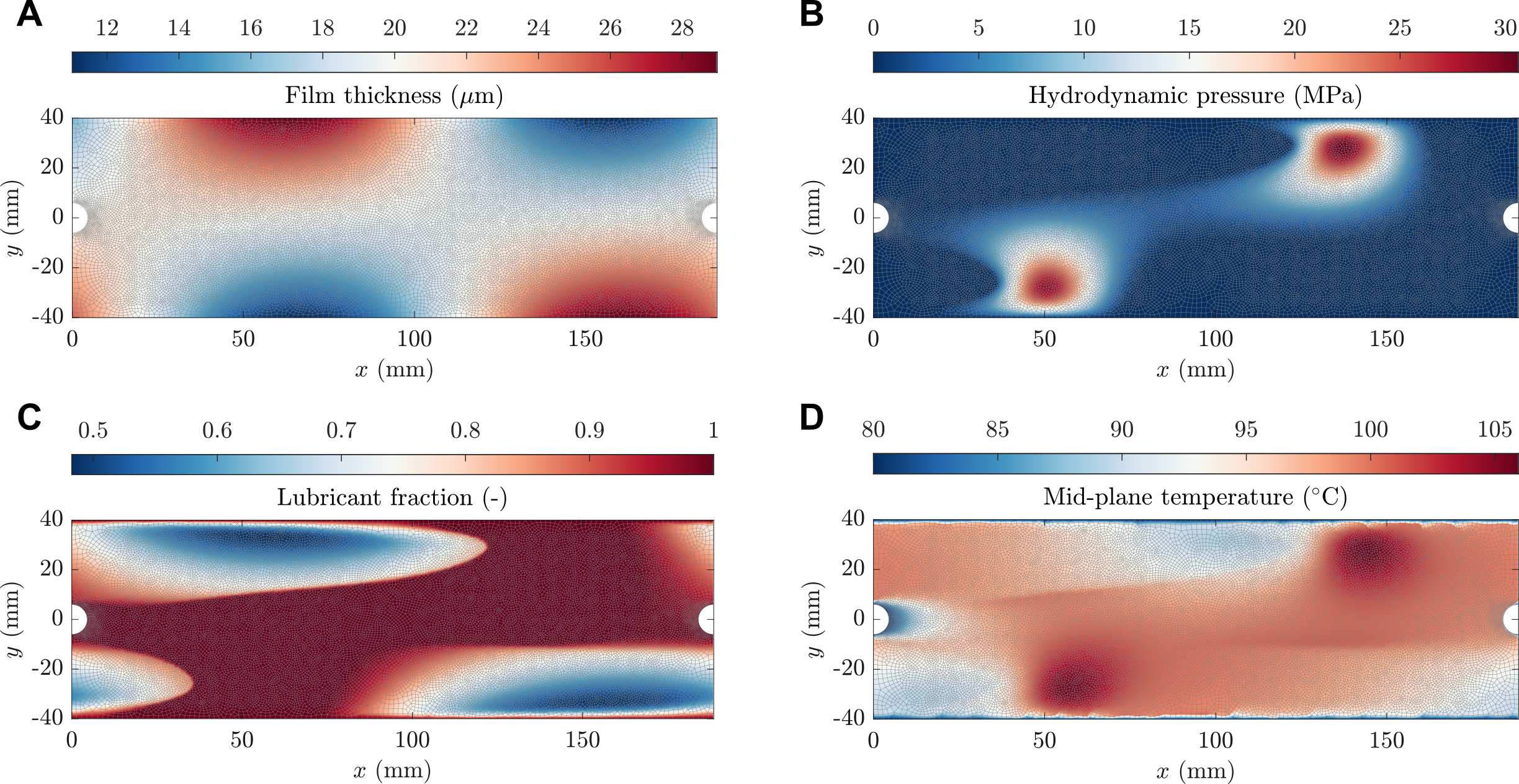}
   \caption{Thermo-hydrodynamic behaviour of the non-textured journal bearing configuration. (\textbf{A}) Film thickness distribution. (\textbf{B}) Hydrodynamic pressure field. (\textbf{C}) Lubricant fraction distribution. (\textbf{D}) Mid-plane film temperature map.}   
   \label{fig:JBR_NoTexture}
\end{figure}

\autoref{fig:JBR_NoTexture} presents the thermo-hydrodynamic behaviour of the smooth (non-textured) journal bearing configuration used as a baseline reference for comparison with the micro-textured interfaces analysed in this study, providing a quantitative evaluation of the effects of surface texturing (dimples, herringbone and sawtooth patterns) on lubricant film geometry, hydroydnamic pressure generation, evolution of cavitated regions and heat dissipation.

\newpage
\bibliography{mybibfile}

\end{document}